\documentclass[onecolumn]{aa} 
\usepackage{graphicx} 
\usepackage{soul} 
\usepackage{txfonts}
\usepackage{bm}
\usepackage{longtable}
\usepackage{supertabular}
\usepackage{xcolor}
\usepackage[normalem]{ulem}\subtitle{}
 \author{
         Domenico Logoteta\inst{1, 2}\fnmsep\thanks{email: domenico.logoteta@unipi.it}
          \and          
         Albino Perego\inst{3,4}\fnmsep\thanks{email: albino.perego@unitn.it}
          \and 
         Ignazio Bombaci\inst{1, 2}\fnmsep\thanks{email: ignazio.bombaci@unipi.it} 
                   }    
 \institute {Dipartimento di Fisica ``E. Fermi'',       
            Universit\`a di Pisa, Largo B. Pontecorvo, 3 I-56127 Pisa, Italy
             \and
            INFN, Sezione di Pisa, Largo B. Pontecorvo, 3 I-56127 Pisa, Italy
             \and 
            Dipartimento di Fisica, Universit\'a di Trento, Via Sommarive 14, 38123 Trento, Italy
             \and
            INFN-TIFPA, Trento Institute for Fundamental Physics and Applications, via Sommarive 14, I-38123 Trento, Italy
          }
\date{Received .........;  accepted ......}
\abstract 
{A precise understanding of the equation of state (EOS) of dense and hot matter is key to modeling relativistic astrophysical environments, including core-collapse supernovae (CCSNe),
protoneutron star (PNSs) evolution, and compact binary  mergers.}
{
In this paper, we extend the microscopic zero-temperature BL (Bombaci and Logoteta)  
nuclear EOS 
to finite temperature 
and arbitrary nuclear composition.
We employ this new EOS to describe hot $\beta$-stable nuclear matter and to compute various structural properties of nonrotating PNS. 
We also apply the EOS to perform dynamical simulations of a spherically symmetric CCSN.} 
{The EOS is derived using the finite temperature extension of the Brueckner--Bethe--Goldstone quantum many-body theory  
in the Brueckner--Hartree--Fock approximation. 
Neutron star properties are computed by solving the Tolman--Oppenheimer--Volkoff structure equations numerically. 
The sperically symmetric CCSN simulations are performed using the AGILE-IDSA code.} 
{Our EOS models are able to reproduce typical features of both PNS and spherically symmetric CCSN simulations.
In addition, our EOS model is consistent with present measured neutron star masses and particularly with the 
masses: $M = 2.01 \pm 0.04 \, M_{\odot}$ and $M = 2.14^{+0.20}_{-0.18} \, M_{\odot}$ of the neutron stars in PSR~J0348+0432 and PSR J0740+6620 respectively. 
Finally, we suggest a feasible mechanism to produce low-mass black holes ($M \sim 2M_{\odot}$) that could have far-reaching consequences for interpreting the gravitational wave event GW190814  as a BH--BH merger.} 
{}
\authorrunning{Logoteta et al.}

\titlerunning{Microscopic equation of state of hot nuclear matter
}

\begin{document}

\title{Microscopic equation of state of hot nuclear matter for numerical relativity simulations}

\maketitle

\section{Introduction} 

One of the main inputs for the study of high-energy astrophysical systems, like core-collapse supernovae \citep[CCSNe; see, e.g.,][for recent reviews]{burrows_review_2013,janka_review_2016}, protoneutron stars \citep [PNSs; e.g.,][]{prak97,roberts_PNS_2017}, and binary compact stars mergers including at least one neutron star \citep[e.g.,][]{baiotti17,Shibata_review_2019,Radice_review_2020}, is the relationship between matter pressure ($P$), energy density ($\varepsilon$), baryonic density ($n$), and temperature ($T$). 
The relationships between these physical quantities,  together with information on the matter composition, make up the equation of state (EOS) of the system  \citep{prak97,Latt-Prak-2016,oertel2017}.  
In recent years, increasing interest surrounding the above-mentioned astrophysical phenomena  has stimulated several studies that aimed to provide as accurate a description as  possible of matter under extreme conditions of density and temperature.
Because of the huge density and temperature variations expected during the dynamical evolution of collapsing stellar cores and merging neutron stars \citep{Fischer_2014,Hanauske_2018,perego19}, 
the derivation of an EOS suitable for their description requires an enormous theoretical and numerical effort. 
Such an EOS needs to  be capable of describing matter in a wide range of conditions, ranging from homogeneous nuclear matter down to a mixture of fully ionized atoms.
To this end, several finite-temperature, composition-dependent EOSs have been developed according to different frameworks and philosophies; among the many, we only mention EOS models based on nonrelativistic Skyrme interactions \citep{LS, SRO}, variational approaches \citep{togashi17}, and relativistic mean field (RMF) models \citep{TM1,SFHO,DD2,IUFSU}.

The recent detection of gravitational waves (GWs) produced in  binary neutron star mergers \citep[BNSMs; see, e.g.,] []{gw5, gw6} has provided a strong boost to the research on dense matter physics.
Gravitational wave astronomy has thus opened up the possibility to explore matter under extreme conditions in regimes that would be inaccessible by experiments on earth.
Gravitational waves from the inspiral phase of BNSMs \citep{gw170817_eos}, especially when combined with robust lower limits on 
the maximum mass of neutron stars (NSs), can set stringent constraints on the EOS of cold (i.e., $T = 0$) nuclear matter 
up to two to three times nuclear saturation density $n_0 = 0.16~\mathrm{fm^{-3}}$ 
(corresponding to $\sim 2.6 \times 10^{14}\mathrm{g~cm^{-3}}$).
The post-merger GW signal and the associated electromagnetic signature 
could provide unique information on the EOS of hot ($T \ge 10 \, \mathrm{MeV}$) and dense ($n \ge 3 n_0$) stellar matter \citep[see e.g.,][]{Radice2018,Radice2018b}. 
In particular, post-merger GWs could carry information about the possible appearance of new particle species (e.g., hyperons) or new phases of matter \citep[e.g., deconfined quark matter; see e.g.,][]{shibata2005,BJ2012,Hotokezaka2013,Bernuzzi2015,RT2016,Bernuzzi2016,Zappa2018,Bauswein2019,Most2020}. In addition, the combined mass-radius measurements provided by Neutron Star Interior Composition Explorer  \citep[NICER; see, e.g.,][] {nicer1, nicer2} set other independent, robust constraints on the high-density EOS.

The purpose of the present work is to extend the microscopic zero-temperature EOS of nucleonic matter derived by \cite{BL} to finite temperature. 
Here "microscopic" refers to the fact that the EOS has been derived within a specified quantum many-body approach starting from two-body and three-body nuclear interactions which reproduce nucleon-nucleon (NN) scattering data, the experimental binding energy of light ($A = 3, 4$) atomic nuclei, and the empirical saturation properties of nuclear matter. 

Various zero-temperature microscopic EOS models are available in the literature \citep[e.g.,][]{wff88, bbb97, apr98}  
and have mainly been employed to describe neutron star structure.  
Some of these EOSs have also been used in BNSM simulations adding a thermal contribution from the so-called 
$\Gamma$-law prescription \citep{BJO2010, kiuchi14}.   

The nuclear many-body problem at finite temperature has been considered by several authors and various  techniques have been developed to handle this very complicated topic. 
Among the most popular ones are those based on Green's function method \citep{FM03,FM05,rios06}, on 
Bloch-De Dominicis (BD) diagrammatic expansion \citep{BD58,BD59,BD59_1}, and, 
more recently, on many-body perturbation theory \citep{fiorilla12}. 
Thus several microscopic EOSs have been extended to finite temperature 
\citep[e.g.,][]{bl94,BF99,zuo_2004,burgio2011,WHK_2015,drisch_2016,CPR_2018,carbone-schwenk_2019}.  

In the present work, as in \citet{BL}, we model the neutron star core as a uniform charge-neutral fluid of neutrons, protons, electrons, and muons. 
The EOS is calculated for arbitrary lepton fractions,  
densities, and temperatures as is indeed required by CCSNe and BNSM simulations in order to describe the behavior of matter which, in the more general case, may be out of equilibrium with respect to the weak interactions. 
We note that even considering this ``simplified'' picture which does not take into account "exotic" degrees of freedom (like hyperons or deconfined quarks), the determination of the EOS from the underlying nuclear interactions 
is still a very challenging theoretical problem.  
In fact, it is necessary to calculate the EOS under extreme conditions of density, neutron-proton asymmetry, and temperature in a regime where the EOS is poorly constrained by nuclear data and experiments.  

In order to describe the interactions between nucleons, consistently with \cite{BL}, we use the version $b$ 
of the local chiral two-nucleon potential reported by \citet[][see Table 2 of that  work]{maria_local}, 
supplemented with the three-nucleon force discussed by \cite{logoteta16b}.  
Finally, using the Brueckner--Bethe--Goldstone  \citep[BBG; see, e.g.,][]{bbg1} many-body theory within the Brueckner--Hartree--Fock (BHF) approximation extended to finite temperature, we calculate the 
Helmholtz free energy of asymmetric nuclear matter. All the other relevant thermodynamic quantities 
(pressure, energy density, etc.) can then be obtained using standard thermodynamic relations.    
This finite temperature EOS will be provided in parametric form, ready for use in numerical 
simulations of BNSMs and CCSNe.
In the particular case of $\beta$-stable nuclear matter at finite temperature, we use our new EOS to compute various structural properties of nonrotating protoneutron stars considering several different stellar conditions such as isoentropic and isothermal neutron star sequences.   
In addition, we apply this new EOS to perform some spherically symmetric (1D) CCSN simulations.  

We stress that in the present work we are extending the zero-temperature EOS of \cite{BL} to finite temperature using modern two-body and three-body nuclear interactions based on chiral effective field theory, while employing the existing finite-temperature BHF many-body approach.

The paper is organized as follows: 
in section \ref{BHF_FT} we describe the extension of the BBG many-body theory at finite temperature. 
In section \ref{EOS hadronic}, we present the results of our new finite-temperature, microscopic EOS for stellar matter.  
Section \ref{applications} is devoted to describing several astrophysical applications of our new EOS, namely: the construction of the EOS in $\beta$-equilibrium both in the isoentropic 
and the isothermal cases, the calculation protoneutron stars structure using these EOSs, and finally its application to spherically symmetric simulations of CCSNe. 
In the last section, we summarize our results and outline the main conclusions of the present work.        

\section{Brueckner--Bethe--Goldstone many-body theory at finite temperature}
\label{BHF_FT}

The Bloch-De Dominicis diagrammatic expansion of the grand canonical thermodynamic potential \citep{BD58,BD59,BD59_1} 
is significant in that it corresponds, in the zero temperature limit, to the BBG expansion of the ground-state energy \citep{BF99}.    
Thus, BD expansion can be viewed as the natural extension of the BBG many-body theory at finite temperature.    
In particular,  \cite{BF99} showed that the dominant terms in the BD expansion are the ones that 
correspond to the (zero temperature) BBG diagrams, where the temperature is introduced in the occupation numbers only,  
represented by Fermi distribution functions \citep{bkl93,bl94}.  

More specifically, the BBG many-body theory is based on a linked cluster expansion 
(the hole-line expansion) of the energy per nucleon $\widetilde{E} \equiv E/A$ of nuclear matter.   
The various terms of the expansion can be represented by Goldstone diagrams grouped according 
to the number of independent hole lines (i.e., lines representing empty single particle states in the Fermi sea). 
The basic ingredient in this approach is the Brueckner reaction matrix $G$, which sums, in a closed form, 
the infinite series of the so-called ladder diagrams and takes into consideration the short-range 
strongly repulsive part of the nucleon--nucleon (NN) interaction. 

In the general case of asymmetric nuclear matter with neutron number density $n_n$, 
proton number density $n_p$, total nucleon number density $n = n_n + n_p$, and 
isospin asymmetry (asymmetry parameter),   
\begin{equation}
             \beta = \frac{n_n - n_p}{n} \, ,
\label{asympar}
\end{equation}
the reaction matrix depends on the isospin thi{rd} components $\tau$ and $\tau^\prime$ of 
the two interacting nucleons ($\tau,\tau'  = n, p$).   
Thus, there are different $G$-matrices describing the $nn$, $pp$, and $np$ in medium effective interactions.   
These are obtained by solving the generalized Bethe--Goldstone equations 
\begin{equation}
 G_{\tau \tau^\prime}(\omega,T) =  v 
  + v  \sum_{k_a,k_b} 
\frac{\mid \vec{k_a},\vec{k_b} \rangle  Q_{\tau \tau^\prime} \langle \vec{k_a},\vec{k_b} \mid}
     {\omega - \epsilon_{\tau}(k_a,T) - \epsilon_{\tau^\prime}(k_b,T) + i\eta } \, G_{\tau \tau^\prime}(\omega,T) \;,
\label{bg}
\end{equation}
where ${v}$ is the bare NN interaction (or a density-dependent two-body effective interaction where averaged 
three-nucleon forces are introduced) and the quantity $\omega$ is the so-called starting energy. 
In the present work we consider spin unpolarized nuclear matter,  and therefore in equation (\ref{bg}) and in the following equations, we drop the spin indices to simplify the formalism
{\footnote{Spin polarized nuclear matter within the BHF approach has been considered  
by, e.g., \cite{vb02} and  \cite{bomb+06}.}}.   
The operator 
\begin{equation} 
\mid \vec{k_a},\vec{k_b} \rangle\, Q_{\tau \tau^\prime}\,\langle \vec{k_a},\vec{k_b}\mid  ~ \equiv ~
\mid \vec{k_a},\vec{k_b} \rangle \, Q_{\tau \tau^\prime}(\vec{k_a},\vec{k_b}; n, \beta,T)\,  \langle \vec{k_a},\vec{k_b} \mid \, = \, \Big[1-f_\tau(\vec{k_a},\tilde{\mu}_\tau,T)\Big] \; \Big[1-f_{\tau{^\prime}}(\vec{k_b},\tilde{\mu}_\tau,T)\Big]
\label{pauli}
\end{equation}
is the finite temperature Pauli operator, which can be written in terms of the Fermi distribution function 
\begin{equation}
f_\tau(k,\tilde{\mu}_\tau,T)=\frac{1}{1+{\rm exp}[(\epsilon_\tau(k,T)-\tilde{\mu}_\tau)/T]} 
\end{equation}
of the two nucleons in the intermediate scattering states. 
In the above expression, $\tilde{\mu}_\tau$ is the so-called  auxiliary chemical potential 
and is obtained 
using the normalization condition 
\begin{equation}
n_\tau=\sum_{k} f_\tau(k,\tilde{\mu}_\tau,T)\,, 
\label{mu-tilde}
\end{equation}
where $n_\tau$ is the number density of neutrons ($\tau = n$) or  protons ($\tau = p$).  
The Bethe--Goldstone equation therefore describes the two nucleons scattering in the presence of other nucleons, 
and the Brueckner $G$-matrix represents the effective interaction between two nucleons in the nuclear medium and properly takes into account the short-range correlations arising from the strongly repulsive core in the bare NN interaction.  

The single-particle energy $\epsilon_\tau(k,T)$ of a nucleon ($\tau = n, p$) with momentum $\vec{k}$ and mass $m_\tau$ is given by
\begin{equation}
       \epsilon_{\tau}(k,T) = \frac{\hbar^2k^2}{2m_{\tau}} + U_{\tau}(k,T) \ ,
\label{spe}
\end{equation}
where $U_{\tau}(k,T)$ is a single-particle potential that represents the mean field felt by a nucleon 
due to its interaction with the surrounding nucleons in the medium. 
In the BHF approximation of the BBG theory, $U_{\tau}(k,T)$
is calculated through the real part of the on-energy-shell $G$-matrix \citep{BBP63,HM72} 
and is given by: 
\begin{equation}
U_{\tau}(k,T) = \sum_{\tau^\prime} \sum_{k'} 
               \mbox{Re} \ \langle \vec{k},\vec{k^\prime} \mid 
               G_{\tau\tau^\prime}(\omega^*,T) \mid \vec{k},\vec{k^\prime} \rangle_a \ f_{\tau^\prime}(\vec{k^\prime},\tilde{\mu}_\tau,T) \;,
\label{spp}
\end{equation}
where $\omega^* = \epsilon_{\tau}(k,T) + \epsilon_{\tau'}(k',T),$ and the matrix elements are properly antisymmetrized. 
We make use of the so-called continuous choice \citep{jeuk+67,gra87,baldo+90,baldo+91} for 
the single-particle potential $U_{\tau}(k,T)$ when solving the Bethe--Goldstone equation.    
As shown by \cite{song98} and \cite{baldo00}, the contribution of the three-hole-line diagrams to the energy per nucleon $E/A$ is minimized in 
this prescription and a faster convergence of the hole-line expansion for $E/A$ is achieved with respect to the  
gap choice for $U_{\tau}(k,T)$.  

We finally performed the usual angular average of the Pauli operator and of the energy denominator  \citep{gra87,baldo+91} in the 
Bethe--Goldstone equation (\ref{bg}).  
We note that, as opposed to the zero-temperature case, the average of the Pauli operator should be performed numerically at finite temperature. 
In this way, the Bethe-Goldstone equation can be expanded in partial waves.   
In the present work, we consider partial wave contributions up to a total two-body angular momentum $J_{\rm max} = 8$.  
We verified that the inclusion of partial waves with $J_{\rm max} > 8$ does not appreciably change our results. 

In this scheme, Eqs.\ (\ref{bg})--(\ref{spp}) have to be solved self-consistently using an iterative numerical procedure. 
Once a self-consistent solution is achieved, the energy per nucleon of asymmetric nuclear matter is  
given by: 
\begin{eqnarray}
\widetilde{E}(n,\beta,T) = 
\frac{E}{A}(n,\beta,T) &=&  
\frac{1}{A}  \sum_{\tau}\sum_{k} \left[ \frac{\hbar^2k^2}{2m_{\tau}}
+\frac{1}{2}  U_{\tau}(k,T)  \right] f_\tau(k,\tilde{\mu}_\tau,T) \;.
\label{energy}
\end{eqnarray} 
The total entropy $S$ has been calculated in the approximation of a mixture of two ideal Fermi gases of quasi-particles in the mean fields $U_{\tau}(k;n, \beta, T)$ \citep{bkl93}: 
\begin{eqnarray}
S(n,\beta,T) &=&    
 - k_B \sum_\tau \sum_k \bigg[f_\tau(k,\tilde{\mu}_\tau,T)\ {\rm ln}\, \Big(f_\tau(k,\tilde{\mu}_\tau,T)\Big) + 
     \Big(1-f_\tau(k,\tilde{\mu}_\tau,T)\Big)\ {\rm ln}\,\Big(1-f_\tau(k,\tilde{\mu}_\tau,T)\Big) \bigg]
\label{entro} \, .
\end{eqnarray} 

Subsequently, the Helmholtz fee energy $F$ can be calculated using the standard thermodynamic relation:
\begin{eqnarray}
   F = E - TS \, .
\label{free1} 
\end{eqnarray} 

In this scheme, the bare two-body nuclear interaction ${v}$ is the only physical input we need for the numerical solution of the Bethe-Goldstone equation. 
As already mentioned, in the present work, as two-body nuclear interaction, we employ the model $b$ of the local chiral N3LO$\Delta$ interaction derived by \citet{maria_local}. 

It is however well established that within the most advanced nonrelativistic quantum many-body approaches  
it is not possible to reproduce the empirical saturation point of symmetric nuclear matter (SNM) 
when using two-body nuclear interactions only. 
The saturation points obtained using different NN potentials lie within a narrow band called the {Coester band} \citep{coester70,day81}, 
with either an overly high saturation density or an overly low binding energy ($B = -E/A$) compared to the empirical values.   
In particular, SNM turns out to be overbound with an overly high saturation density when using modern 
high-precision NN potentials, fitting NN scattering data up to an energy 
of $350$ MeV, with a $\chi^2$ per datum close to $1$ \citep{ZHLi06}. For the nuclear matter 
case, as in the case of few-nucleon systems\ \citep{kalantar12,hammer13,binder16}, three-nucleon-forces (TNFs) are considered to be the missing physical effect of the whole 
picture.    
The inclusion of TNFs is therefore required in order to reproduce a realistic saturation 
point \citep{FP81,bbb97,apr98,Li2008,zuo14,kievsky18}. 
In addition, TNFs are crucial in the case of dense $\beta$-stable nuclear matter to obtain a stiff 
EOS \citep{bbb97,apr98,li-schu_08,chamel11} compatible with the measured masses 
{\footnote{Stellar masses are given in units of  solar mass $M_\odot = 1.989 \times 10^{33}$~g}\,}      
$M = 1.97 \pm 0.04 \, M_\odot$  \citep{demo2010},  
$M = 2.01 \pm 0.04 \, M_\odot$  \citep{anto2013}, and 
$M = 2.14^{+0.20}_{-0.18} \, M_\odot$  \citep{croma19} 
of the neutron stars in PSR~J1614-2230, PSR~J0348+0432, and PSR J0740+6620 respectively. 
 
In the present work we use the same TNFs adopted by \citet{BL} namely the N2LO$\Delta$. 
The parameters of such force were determined in \citet{logoteta16b} in order to reproduce good saturation properties of SNM. The specific parametrization that we use in the present work is the N2LO$\Delta1$ \citep{logoteta16b,BL}.  
As in \citet{logoteta16b} and in \citet{BL}, we derive an effective density-dependent two-body force $v_{NN}^{\rm eff}(n)$   
 starting from the original three-body force by averaging over one of the three nucleons \citep{holt}.      
The Bethe--Goldstone equation (\ref{bg}) is then solved adding this effective density-dependent 
two-body force to the bare NN interaction. This strategy allows us to avoid the need to solve three-body Faddeev 
equations in the nuclear medium (Bethe--Faddeev equations) \citep{bethe65,rajaraman-bethe67} which are currently too complicated to deal with. 

We note that adopting the two- and three-nucleon interactions described above, the resulting EOS is consistent 
with the one derived in \cite{BL} at zero temperature.    

Finally, we want to emphasize that the compatibility between the BBG many-body method and other two independent many-body methods, namely the Fermi hypernetted chain/single-operator chain, and the auxiliary-field diffusion Monte Carlo, was recently checked in benchmark calculations \citep{benchmark2020} of the energy per particle of 
pure neutron matter.   

\section{Equation of state of hot nuclear matter } 
\label{EOS hadronic}

\begin{figure}[t]
\centering
\includegraphics[width = 8.cm]{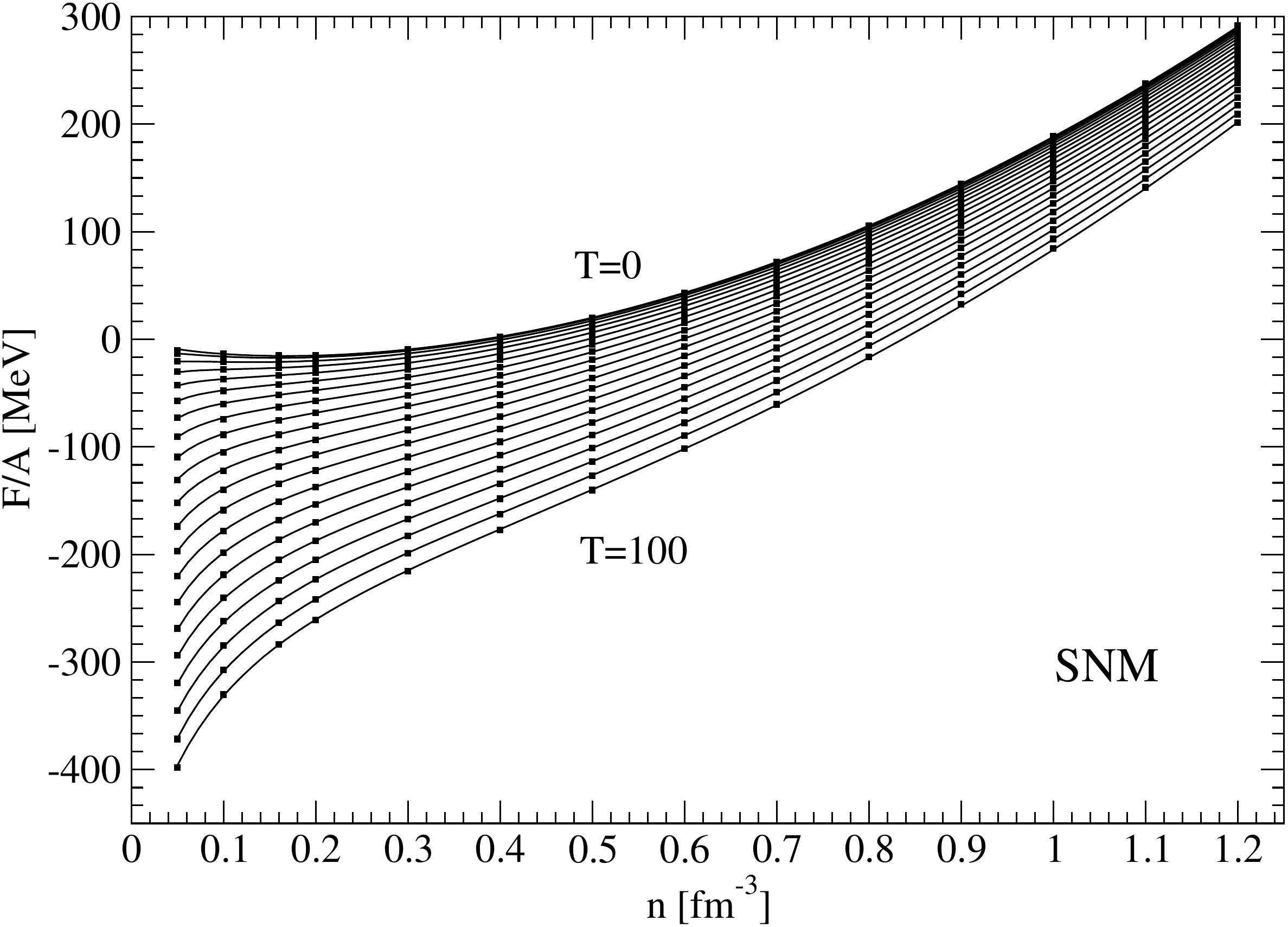} 
\caption{Free energy per particle ($F/A$) of SNM  as a function of the nucleon number density ($n$) for the interaction model considered in this work. The square symbols represent the results of our microscopic BHF calculations, 
whereas the continuous lines represent the energy per particle obtained using the parametrization given by 
eq.~(\ref{free}).  Temperatures (T) in the figure are in {\rm MeV}. }
\label{fig2}
\end{figure}

\bigskip

\begin{figure}[t]
\centering
\includegraphics[width = 8.cm]{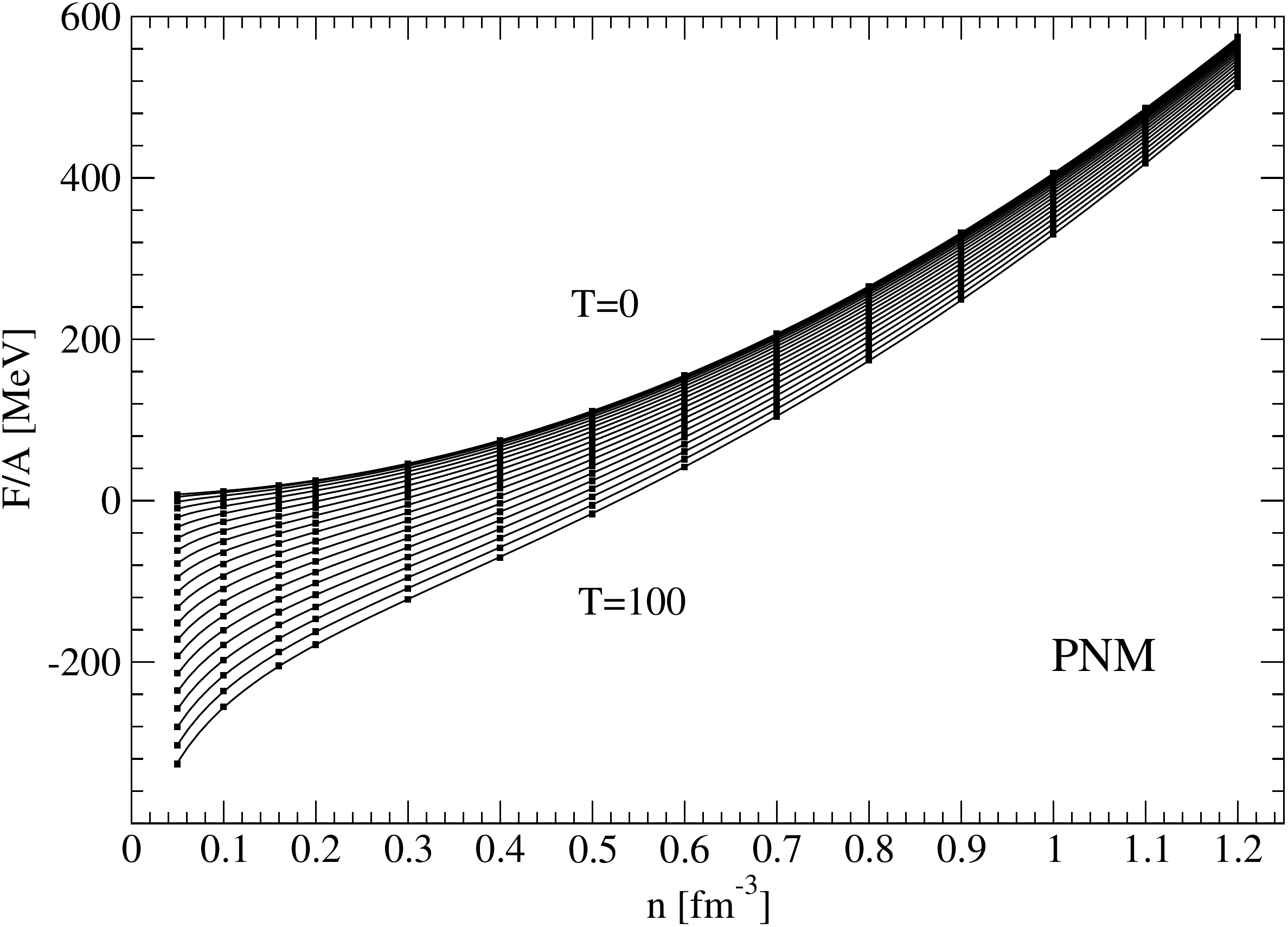} 
\caption{Same as in Figure~\ref{fig2}, but for PNM.}
\label{fig1}
\end{figure}

For baryon densities greater than about $0.05$ ${\rm fm^{-3}}$, the stellar matter can be described as an homogeneous fluid of neutrons and protons (nuclear matter) with leptons (electrons and muons) to guarantee electric charge neutrality. In this density region, the finite temperature BHF approach outlined in the previous section is appropriate to describe the system.  
For densities smaller than about $0.05$ ${\rm fm^{-3}}$, nucleons start to form clusters and, as a consequence, the stellar matter becomes inhomogeneous\footnote{We note however that this consideration strongly depends on the temperature of the system: above some critical temperature ($\sim 18$ {\rm MeV}), at all densities clusters disappear and matter becomes homogeneous.}. 
In these conditions, the BHF approach does not provide a reliable description of the nuclear component and another theoretical framework is therefore required to compute the EOS. We return to this point below.  
 
In the density range $0.05~\mathrm{fm^{-3}} \lesssim  n \lesssim 1.2~\mathrm{fm^{-3}}$, the free energy per nucleon 
$\widetilde{F} = F/A$ of asymmetric nuclear matter can be calculated by solving 
Eqs.\ (\ref{bg})--(\ref{entro}) 
numerically for various values of the asymmetry parameter ($0 \leq \beta \leq 1$) and temperatures 
($0 \lesssim T \lesssim 100~\mathrm{MeV}$).
In order to evaluate the free-energy contribution of asymmetric nuclear matter, we employ the so-called {\it parabolic approximation} in the asymmetry parameter $\beta$ \citep{bl91,bkl93,bl94,WHK16}:    
\begin{equation}
\widetilde{F}(n,\beta,T) = \widetilde{F}(n,\beta=0,T) + F_{\rm sym}(n,T)\, \beta^2 \, ,
\label{F_pa}
\end{equation}
where $F_{sym}(n,T)$ is the so-called nuclear symmetry free energy which, in the $T = 0$ limit, gives 
the nuclear symmetry energy. 
In the parabolic approximation of Eq.\ (\ref{F_pa}), the symmetry free energy can therefore be obtained as the difference between  $\widetilde{F}$ of pure neutron matter (PNM, $\beta = 1$) and that of SNM ($\beta = 0)$: 
\begin{equation}
F_{\rm sym}(n,T)  = \widetilde{F}(n,\beta = 1, T) - \widetilde{F}(n,\beta = 0, T) \, .
\label{Fsym}
\end{equation}
It has been shown that such approximation provides accurate results both at zero \citep{bl91} and finite temperature \citep{bkl93,bl94,WHK16}. 
Therefore, the two basic quantities appearing in Eqs. (\ref{F_pa}) and (\ref{Fsym}) are the free energies per nucleon 
of SNM and of PNM as a function of density and temperature. Their behaviors are shown in Fig.\ \ref{fig2} and\ Fig. \ref{fig1} for SNM and PNM, respectively.     
The filled square symbols represent the results of our microscopic BHF calculations described in the previous section, whereas the continuous lines are the results of the fit obtained using the following parametrization:   
\begin{equation} 
   \widetilde{F}(n,T) \equiv \frac{F}{A}(n,T)=    (a_{0} + a_1 t^2 + a_{2} t^3 + a_{3} t^4 ) \ {\ln}~ \Bigg(\frac{n}{\mathrm{fm}^{-3}}\Bigg) + 
      ( a_4 + a_5 t^2 ) \ n + 
      (a_6 + a_7 t^2 ) \ n^{a_8} +
      \frac{1}{n}(a_9 t^2 + a_{10} t^{a_{11}}) 
      + a_{12} \;,
\label{free}
\end{equation}
where we  define $t \equiv T/(100~\mathrm{MeV})$. The fit parameters $a_i$ for SNM and PNM are reported in 
Table \ \ref{tab_coeff}. 
The two-dimensional fit in the variables ($n,T$) produces a very good $\chi^2/{\rm datum}$ $\sim 1$ for both 
SNM and PNM.
\begin{table*}  
\caption{Value of the coefficients $a_i$ (with $i = 0, 1,...,12$) in the parametrization of the free energy per particle $F/A$ (see Eq.\ \ref{free}) for SNM (first row) and PNM (second row).  
The coefficients $a_0$, $a_1$, $a_2$, $a_3$ and $a_{12}$ are given in MeV, the coefficients $a_4$ and $a_5$ are given in MeV~fm$^{3}$, $a_6$ and $a_7$ in MeV~fm$^{3a_8}$,  
$a_9$ and $a_{10}$  in MeV~fm$^{-3}$, $a_8$ and $a_{11}$ are dimensionless quantities.  
The $\chi^2/{\rm datum}$ resulting from the fit is 1.05 for SNM and 1.1 for PNM.}    
\label{tab_coeff}
\centering
\renewcommand{\arraystretch}{2}
\begin{tabular}{cccccccccccccc}
\hline
\hline
 {~} & $a_0$ & $a_1$& $a_2$& $a_3$& $a_4$& $a_5$& $a_6$& $a_7$& $a_8$& $a_9$& $a_{10}$& $a_{11}$& $a_{12}$ \\
\hline 
SNM &-2.33 & 206.30 & -87.49 & 20.54 & -111.36 & -172.48 & 310.13 & 67.20 & 1.84 & 13.35 & -11.42 & 1.49 & -12.37  \\
\hline
PNM & 3.205 &  158.61 &    -54.64 &   12.11    &  -28.17  & -123.14 &   415.96  &  47.55 &   1.91 &   19.43    &  -18.37 &  1.66 &   17.00  \\
\hline
\hline 
\end{tabular}
\end{table*}

Using these parametrizations of the free energy per nucleon we can easily obtain analytic expressions for all the relevant thermodynamic quantities defining the EOS of hot asymmetric nuclear matter.
To this purpose we introduce the free-energy density as $\mathfrak{f}_N = F/V = n~(F/A)$, 
and then the neutron ($\mu_{n}$) and proton ($\mu_{p}$) chemical potentials can be obtained using:
\begin{equation}
\mu_{n}\,(n,\beta,T) = \frac{\partial \mathfrak{f}_N}{\partial n_n}\bigg\vert_{T,V,n_p} \, ,
\quad\quad
\mu_{p}\,(n,\beta,T) = \frac{\partial \mathfrak{f}_N}{\partial n_p}\bigg\vert_{T,V,n_n} \,,
\label{chempot}
\end{equation}
and the entropy density $s_{N} = S/V = n~(S/A)$ (in units of Boltzmann constant $k_B$) 
can be obtained using:
\begin{equation}
s_N(n,\beta,T) =  - \frac{\partial \mathfrak{f}_N}{\partial T}\bigg\vert_{V,n_n,n_p} \,.
\label{entroh}
\end{equation}  
The nuclear pressure is then given by:
\begin{equation}
P_N(n, \beta, T)  =  \mu_n n_n + \mu_p n_p - \mathfrak{f}_N \, ,
\end{equation}
and finally the energy density by:
\begin{equation}
\varepsilon_N = \mathfrak{f}_N + T s_N\,. 
\label{energydens}
\end{equation}
%
\begin{figure}[t]
\centering
\includegraphics[height = 6.cm]{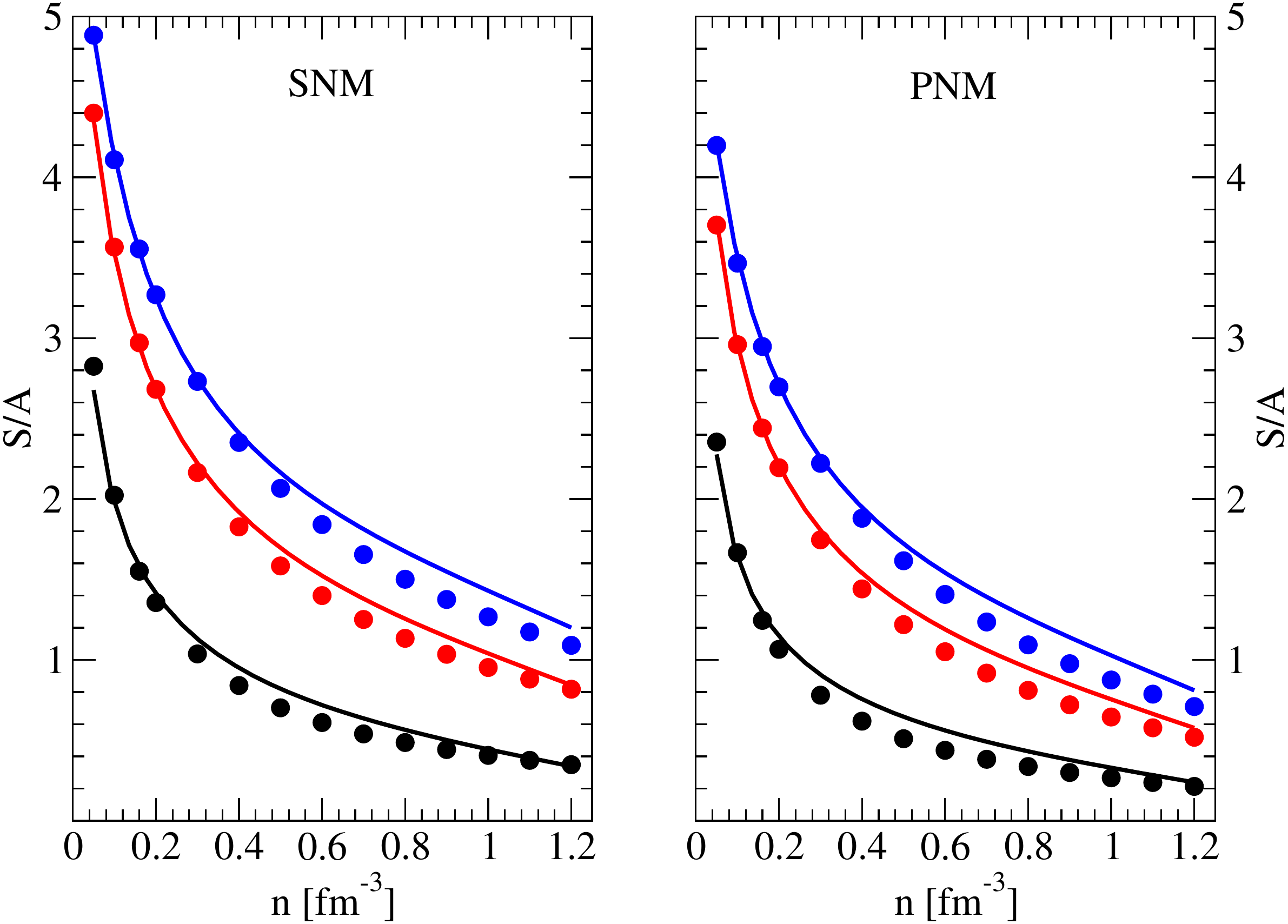} 
\caption{Entropy per nucleon ($S/A$) as a function of the nucleon density $n$ at different temperatures,  
$T = 20, 50, 70$ MeV (from bottom to top) for SNM (left panel) and PNM  (right panel).  
The continuous lines represent $S/A$ calculated using Eq.~(\ref{entroh}) in which we employ the fit formula (Eq.~(\ref{free})) for the free energy. The circular symbols represent the entropy per nucleon, calculated using Eq.~(\ref{entro}) in the approximation of a mixture of two ideal Fermi gases of quasi-particles in the BHF mean fields. } 
\label{fig_check1}
\end{figure}

\bigskip

\begin{figure}[t]
\centering
\includegraphics[height = 6.cm]{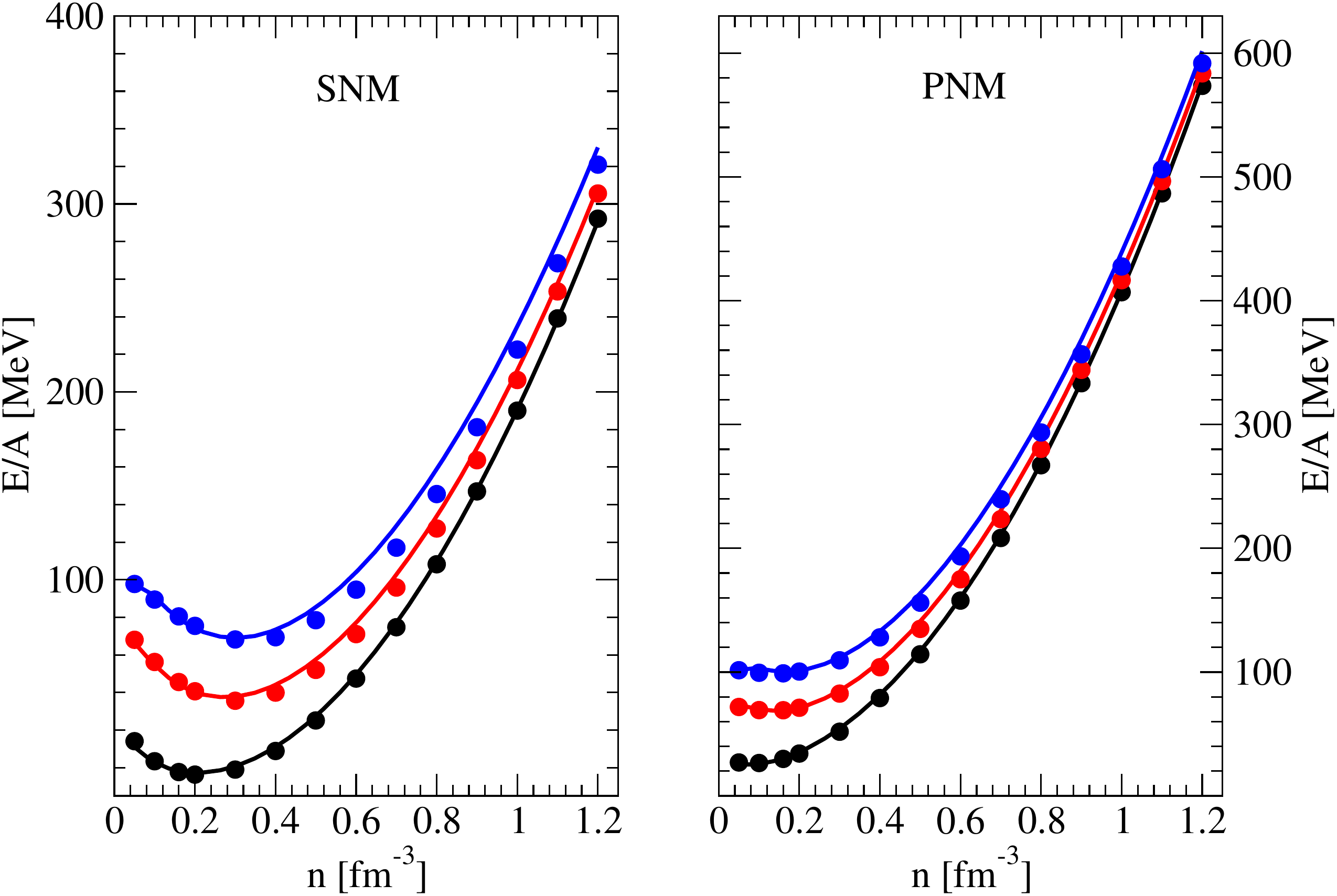} 
\caption{Energy per particle ($E/A$) as a function of the nucleon density $n$ at different temperatures 
$T = 20, 50, 70$ MeV (from bottom to top) for SNM (left panel) and PNM  (right panel).  
The continuous lines represent $E/A =\varepsilon_N/n$ calculated using Eq.~(\ref{energydens}) whereas 
the circular symbols represent the results of the self-consistent BHF calculations, i.e., Eq.~(\ref{energy}). }
\label{fig_check2}
\end{figure}

Before proceeding, we wish to comment on the procedure that we have outlined above. 
It is well known that in the BHF approximation the nucleon chemical potentials, calculated according to the thermodynamic definition in Eq. (\ref{chempot}), differ from the auxiliary chemical potentials in Eq. (\ref{mu-tilde}) 
used in the many-body approach. As a consequence, the Hugenholtz-Van Hove theorem \citep{HVH} is not fulfilled.  
In the case of SNM at zero temperature, the Hugenholtz-Van Hove theorem states that the energy per nucleon at the saturation density must be equal to the Fermi energy of the quasi-particles. 
A violation of this theorem is therefore a signal of the thermodynamic inconsistency of the BHF approximation.  
This problem can be solved by going beyond the BHF approximation
of the single-particle energy by including some extra diagrams of the hole-line expansion of the mass operator  
\citep[see e.g.,][]{HM72} as discussed in \cite{baldo+90,zuo_M2,zuo_FT}. 
In particular, it turns out that the inclusion of the so-called rearrangement contribution $M_2$, which is a second-order diagram in the $G$-matrix, and the renormalization contributions of third  and fourth order 
in the $G$-matrix, strongly improve the thermodynamic consistency of the perturbative many-body approach. 
Inclusion of such diagrams is beyond the scope of the present work.  
Here we employ a different scheme: we use the expression of the free energy per particle 
obtained from our microscopic calculations and provided by Eqs.~(\ref{F_pa})-(\ref{free}), and afterwards 
we derive all the other thermodynamic quantities from it; see Eqs.~(\ref{chempot}) - (\ref{energydens}). 
Within this scheme we obtain a thermodynamically consistent EOS. 
This strategy has been successfully adopted in various works \citep{burgio10,burgio2011} which make use 
of the finite temperature extension of the BHF approximation.  

In order to compare the microscopic BHF calculations 
with respect to the thermodynamically consistent EOS derived 
from the fit of the free energy,  
we show in Fig. \ref{fig_check1} the entropy per nucleon ($S/A$) 
as a function of the nucleon density $n$ at different temperatures ($T =$ 20, 50, 70 MeV) for SNM (left panel) and PNM  (right panel).  
The continuous lines represent $S/A$ calculated using Eq.~(\ref{entroh}) in which we employ the fit formula shown in Eq.~(\ref{free}) for the free energy. 
The circular symbols represent the entropy per nucleon, calculated using Eq.~(\ref{entro}), in the approximation 
of a mixture of two ideal Fermi gases of quasi-particles in the BHF mean fields $U_{\tau}(k;n, \beta, T)$.  
The agreement between the two results is relatively good but deteriorates at the  highest temperatures and densities.  
In Fig.~\ref{fig_check2} we report the same type of analysis as shown in Fig.~\ref{fig_check1} but for the 
energy per particle $E/A$. 
The continuous lines represent $E/A =\varepsilon_N/n$ calculated using Eq.~(\ref{energydens}) whereas 
the circular symbols represent the results of the self-consistent BHF calculations of Eq.~(\ref{energy}). 
The agreement between the two results is particularly  good at all the considered  temperatures and densities. 

We finally note that in the present paper we do not employ any approximation in the calculation of the BHF 
single-particle neutron and proton potentials, which are evaluated at finite temperature by solving  
Eqs.\ (\ref{bg}), (\ref{spe}), and (\ref{spp}) consistently. 
A so-called frozen correlation approximation for $U(k,T)$, namely the use of the values 
calculated at zero temperature, has often  been applied  instead in other works \citep[e.g.,][]{burgio2011}. 

Additionally, we note that nuclear chiral interactions are usually employed up to two to three times the nuclear saturation density $n_0$. As shown for instance in  \citet{drisch19}, \citet{benchmark2020}, the uncertainty estimates on many-body nuclear observables increase for increasing nuclear density. 
However, to describe astrophysical systems like NSs or BNS mergers we need to extend the density range 
of our many-body calculations above the trusted region of chiral effective field theory. 
This scheme is more consistent than  the use of piecewise polytropic extensions of the EOS for example, which are sometimes used in numerical general relativity simulations.       
 
We now consider the low-density region, that is, baryon densities $n \leq 0.05 \ {\rm fm}^{-3}$ where nucleon 
clusters appear. For $T \gtrsim 0.5~{\rm MeV}$, matter is in nuclear statistical equilibrium (NSE), i.e., strong and electromagnetic interactions are in equilibrium. Under these conditions, 
the possible nuclear species are  also  fully determined by the baryon density, temperature, and asymmetry parameter. 
The latter parameter is often expressed in terms of the total \footnote{In this context, "total" refers to both free and bound protons in nuclei.} proton fraction, $Y_p$.
In the present work in order to describe all the thermodynamical quantities for $n \leq 0.05 \ {\rm fm}^{-3}$, we consider the HS EOS {hempel10} in its publicly available tabular forms. This EOS was derived in the framework of the RMF approach and models the appearance of nuclei through an excluded volume approach 
\citep[see][for more details]{hempel10}. In particular, we consider two specific EOS tables derived from two different RMF nuclear interaction parametrizations: the TM1 \citep{TM1} and SFHo \citep{SFHO} ones. 
As detailed in these latter studies, these EOSs take into account matter clusterization in the low-density regions and therefore complement the EOS provided by our microscopic BHF approach. 
Specifically, we performed a linear interpolation in all the relevant thermodynamical quantities between the microscopic BHF and the RMF calculations in the region  $0.05 \ {\rm fm}^{-3} \leq n \leq 0.08 \ {\rm fm}^{-3}$. We checked that the match between the EOSs provided by the two different approaches is smooth enough to avoid the creation of fictitious phase transitions. 
Despite its simplicity, such an interpolation preserves the thermodynamical consistency of the two underlying EOSs. 
In addition, the spherically symmetric CCSN simulations  discussed in the following section serve as additional verification of the construction of the complete EOS. Such simulations are indeed very sensitive in a wide density range: from very low density ($\sim 10^{-13}$ {\rm fm$^{-3}$} ) up to $\sim 0.32$ {\rm fm$^{-3}$}. The closeness of the match between the BHF EOS and the EOSs based on the RMF approach  can therefore be  additionally tested by checking that the typical features of 1D CCSNe simulations are correctly reproduced.            
We finally note that we have also compared our results with others obtained using different EOSs for the low-density region, such as the one reported by \citet{togashi17} which were derived using a variational approach, or one of the Lattimer-Swesty \citep{LS} EOS family (specifically the one with compressibility modulus $K_\infty=220$ {\rm MeV}). We did not find appreciable differences at least for our purposes.


\section{Astrophysical applications}
\label{applications}


In astrophysical applications involving nuclear matter, the astrophysical plasma is assumed to be charge neutral and composed of hadrons (nuclei, nucleons), massive leptons (electron, muons, and their antiparticles), and photons in thermodynamical equilibrium. Hadrons are usually considered in NSE.
Under these assumptions, the most general full EOS could be expressed as a function of baryon density,  temperature, and a couple of particle abundances, such as for example the net (i.e., subtracted by the antiparticle contribution) abundance of electrons, $Y_e \equiv n_e/n = \left( n_{e^-} - n_{e^+} \right)/n$
 and of muons, $Y_{\mu} \equiv n_{\mu}/n = \left( n_{{\mu}^-} - n_{{\mu}^+} \right)/n$. 
The abundances $Y_e$ and $Y_{\mu}$ , together with that of protons ($Y_p$), must satisfy the local charge neutrality.
Depending on the matter conditions and on the explored timescales,  a gas of trapped neutrinos in weak equilibrium can also be present.

In order to obtain a complete EOS suitable for astrophysical applications we have to add the contributions of massive leptons ($L$), photons ($\gamma$), and possibly neutrinos ($\nu$, in three different species) to the hadronic entropy, energy density, and pressure, namely: $S = S_N+S_L+S_{\nu}+S_\gamma$, \,  $\varepsilon=\varepsilon_N+\varepsilon_L+\varepsilon_{\nu}+\varepsilon_\gamma$, and $P=P_N+P_L+P_{\nu}+P_\gamma$. 
The energy density ($\varepsilon_i$), pressure ($P_i$), and entropy ($S_i$) for the massive leptons and neutrinos
($i=L, \ \nu$) have been computed using the expressions for relativistic ideal Fermi gases at finite temperature 
with $m_e c^2$ = 0.511~MeV, $m_\mu c^2 = 105.658$~MeV and $m_\nu=0$  \citep{timmes}.  Additional requirements, including weak equilibrium conditions, are usually needed to fix the relative amounts of electrons, muons, and neutrinos (see below for a few concrete examples). The resulting number densities fix the particle chemical potentials. Photon contributions are instead calculated according to the Bose statistic and result in well-known analytic expressions.
In the remainder of this section, we show results related to the application of our new EOS to various astrophysical systems. We first discuss the calculation of the EOS for stellar $\beta$-stable nuclear matter; we then determine the corresponding structure for nonrotating protoneutron stars, and finally analyze the results of simulations of CCSNe. 
We also note that both the $T=0$ and the finite temperature versions of the BL EOS have already been used in numerical relativity simulations of BNSMs \citep{endrizzi18,Bernuzzi20}.

\subsection{Equation of state for  $\beta$-stable nuclear matter}
\label{Sec:EOS for asymmetric beta-stable nuclear matter}

As a first application of our new EOS, we derive a finite-temperature, $\beta$-stable EOS suitable for describing hot neutron star matter.
The latter is required to model PNSs produced in successful CCSN explosions and the possibly metastable remnant of a BNSM in the case where a black hole (BH) does not form promptly at merger. In both cases, the large temperatures resulting from the collapse or merger dynamics produce a copious amount of neutrinos, which initially form a trapped neutrino gas and diffuse out over the diffusion timescales ($\sim$ seconds). The fate of these hot neutron stars, as well as their properties, depends on the progenitor system. Numerical simulations show that in both scenarios the bulk of matter is characterized by a quasi-uniform, low-entropy-per-baryon profile $\widetilde{S} = S/A \approx 1-3 $ (in units of the Boltzmann constant $k_B$), which decreases only over 
the cooling timescale because of neutrino 
emission \citep{BL86,prak97,pons99,Fischer2010,Hudepohl2010,roberts2012,Kastaun2015,perego19}. 
In the PNS case, the total lepton fraction is initially rather high ($Y_{l_e} \equiv (n_{e} + n_{\nu_e})/n \sim 0.3 - 0.35$, where $n_{\nu_i}$ is the net electron neutrino density) and the cooling process is associated with a significant deleptonization towards cold neutrino-less weak equilibrium.  
In the BNSM case, matter is characterized by an initially low $Y_{l_e} \lesssim 0.1 $, and experiences a leptonization process because of decompression and temperature increase. 

To account for the different NS profiles and cooling phases, we model hot nuclear matter by adopting an isoentropic EOS and fixing the value of the entropy per baryon $S/A$ in the range $0\le  S/A \le 3 $. 
In the following section, we also consider the case of a isothermal EOS although this case is not particularly realistic.
We consider both the case of neutrino-less and neutrino trapped matter. In the latter case we fix the total lepton fractions 
to $Y_{l_e}=0.33$ and$Y_{l_\mu}=0$. 
 
In order to calculate the composition of $\beta$-stable nuclear matter we solve the equations 
for chemical equilibrium 
at a given nuclear density $n$ and temperature $T$:
\begin{equation}
\mu_n-\mu_p=\mu_e-\mu_{\nu_e}\;, \ \ \ \ \ \ \ \mu_\mu = \mu_{e} - \mu_{\nu_e}  + \mu_{\nu_\mu} \;.  
\label{beta1}
\end{equation}
In addition, charge neutrality requires  
\begin{equation}
   n_p = n_e +  n_{\mu} \,. 
\label{beta2}
\end{equation}
In the case of neutrino-free matter we set: $\mu_{\nu_e}=\mu_{\nu_\mu}=0$. 
Once the solutions of the above equations have been determined, all the other thermodynamical quantities can be calculated according to standard relations as discussed in the previous section.

\begin{figure}[t]
\centering
\includegraphics[width = 9.cm]{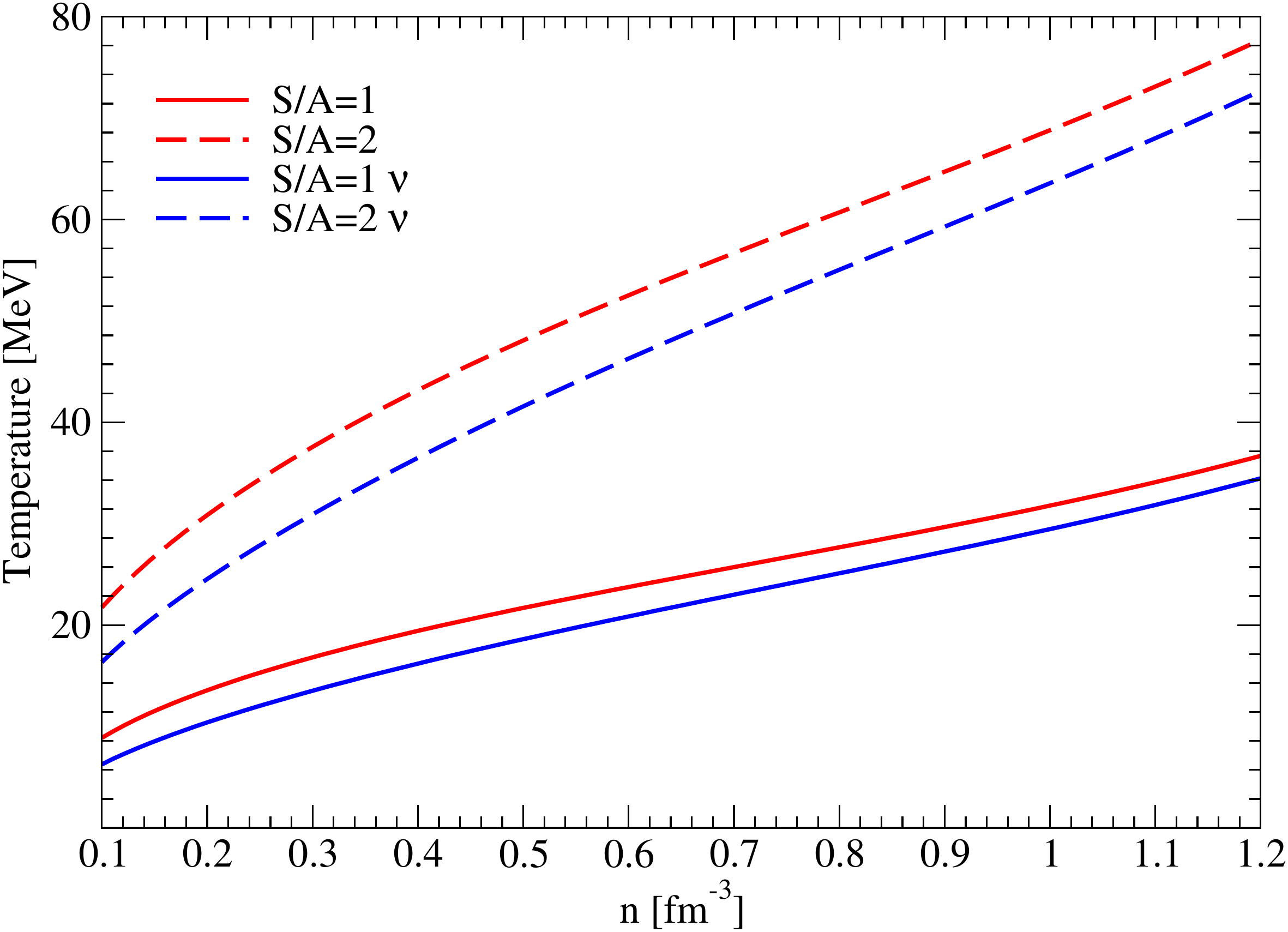} 
\caption{Temperature profiles as a function of nuclear density $n$ for isoentropic ($S/A =$~const)  
$\beta$-stable EOSs in the case of neutrino-free (red lines) and neutrino trapped (blue lines) matter. }
\label{temperature_S}
\end{figure}

In Fig.\ \ref{temperature_S} we report the temperature profiles as a function of the nuclear density for  isoentropic, $\beta$-stable EOSs with constant entropy per baryon  $S/A = 1$ (continuous lines) and $S/A=2$ (dashed lines), in the case of neutrino-free and neutrino-trapped matter. 
As matter is compressed, the temperature rises.  
As expected,  larger temperatures
are reached when increasing the value of the entropy per baryon. For a fixed nuclear density, the effect of neutrino-trapping 
reduces the value of the temperature compared with the neutrino-free case. This effect is well understood noting that when neutrinos are trapped, a new degree of freedom is introduced in the system which therefore becomes more "disordered" causing the total entropy to increase. 
For a fixed value of $S/A$, and a given nuclear density $n$, a lower temperature is therefore required.

\begin{figure}[t]
\centering
\includegraphics[width = 9.cm]{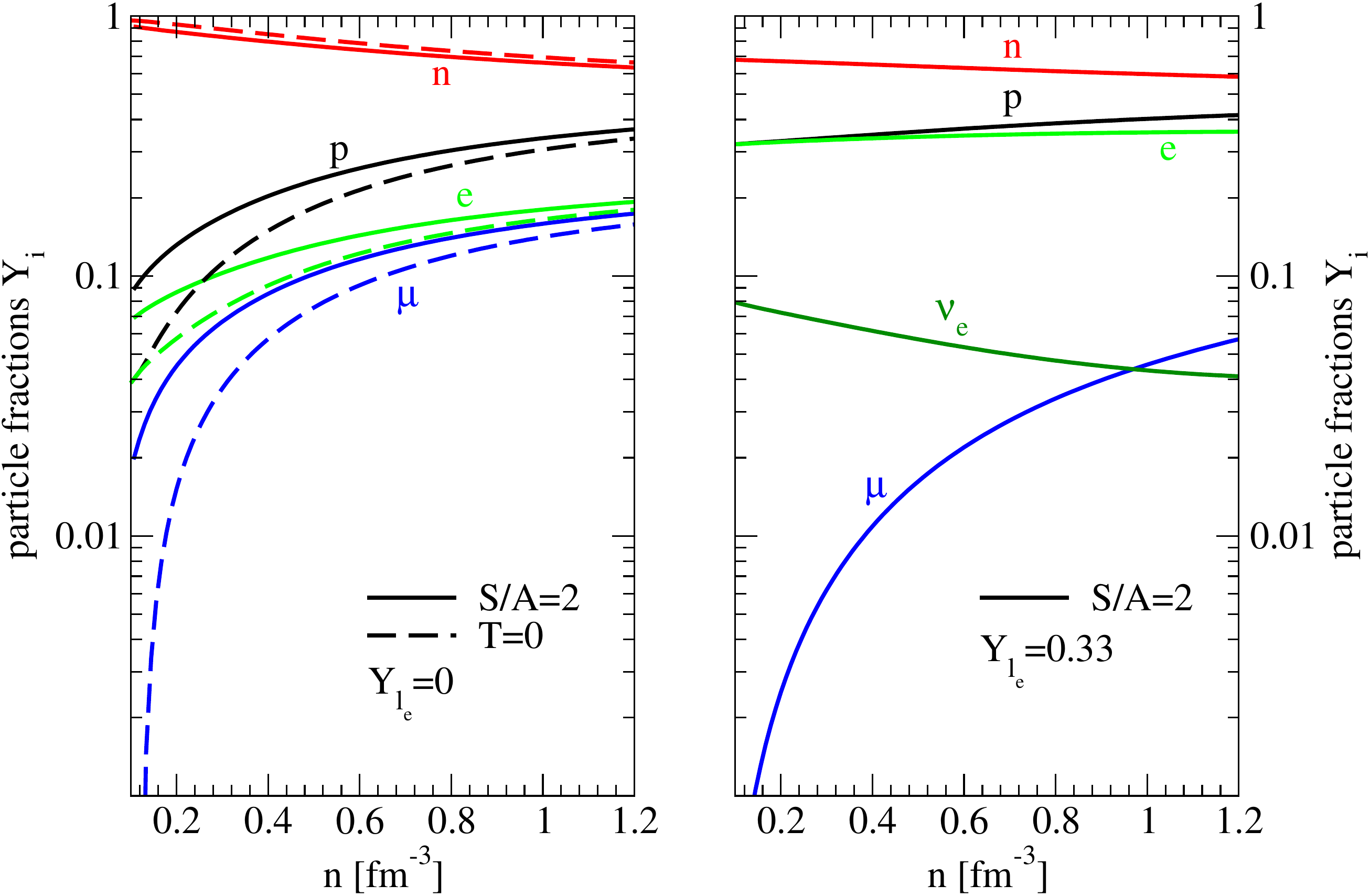} 
\caption{ Composition of dense matter for different physical regimes. In the left (right) panel we show the composition  of $\beta$-stable neutrino-free (trapped) nuclear matter, {i.e.,} the particle fractions $Y_i=n_i/n$ (with $i=n,p,e^-,\mu^-$), in the case of an isoentropic EOS with 
$S/A =2$. In the right panel we fix $Y_{l_e}=0.33$.  
The dashed lines in the left panel represent the composition of the $T=0$ BL EOS \citep{BL} which is reported for comparison.  
See text for further details.  }
\label{composition}
\end{figure}

The composition of isoentropic ($S/A = 2$) $\beta$-stable nuclear matter, {that is,} the particle fractions $Y_i=n_i/n$ (with $i=n,p,e^-,\mu^-$) without (with) neutrino trapping, is shown in Fig.\ \ref{composition}. In the same figure we report the $T=0$ particle composition of the 
BL EOS for comparison \citep{BL}. 
The main effect of thermal contribution is to increase the number of protons and to move the muon onset to smaller density.    
This behavior is similar to that found by many other phenomenological and microscopic approaches \citep{burgio2011}.  We note that in the case of matter including trapped neutrinos (right panel in Fig.\ \ref{composition}), the electron fraction rises and the composition of matter becomes more symmetric in the neutron and proton fractions than in the untrapped case. As a consequence, the EOS of neutrino-trapped  matter is expected to become softer \citep[][see \cite{bomb96} for a detailed discussion on the role of neutrino trapping and the value of the neutron star maximum mass]{prak97}.

\begin{figure}[t]
\centering
\includegraphics[width = 9.cm]{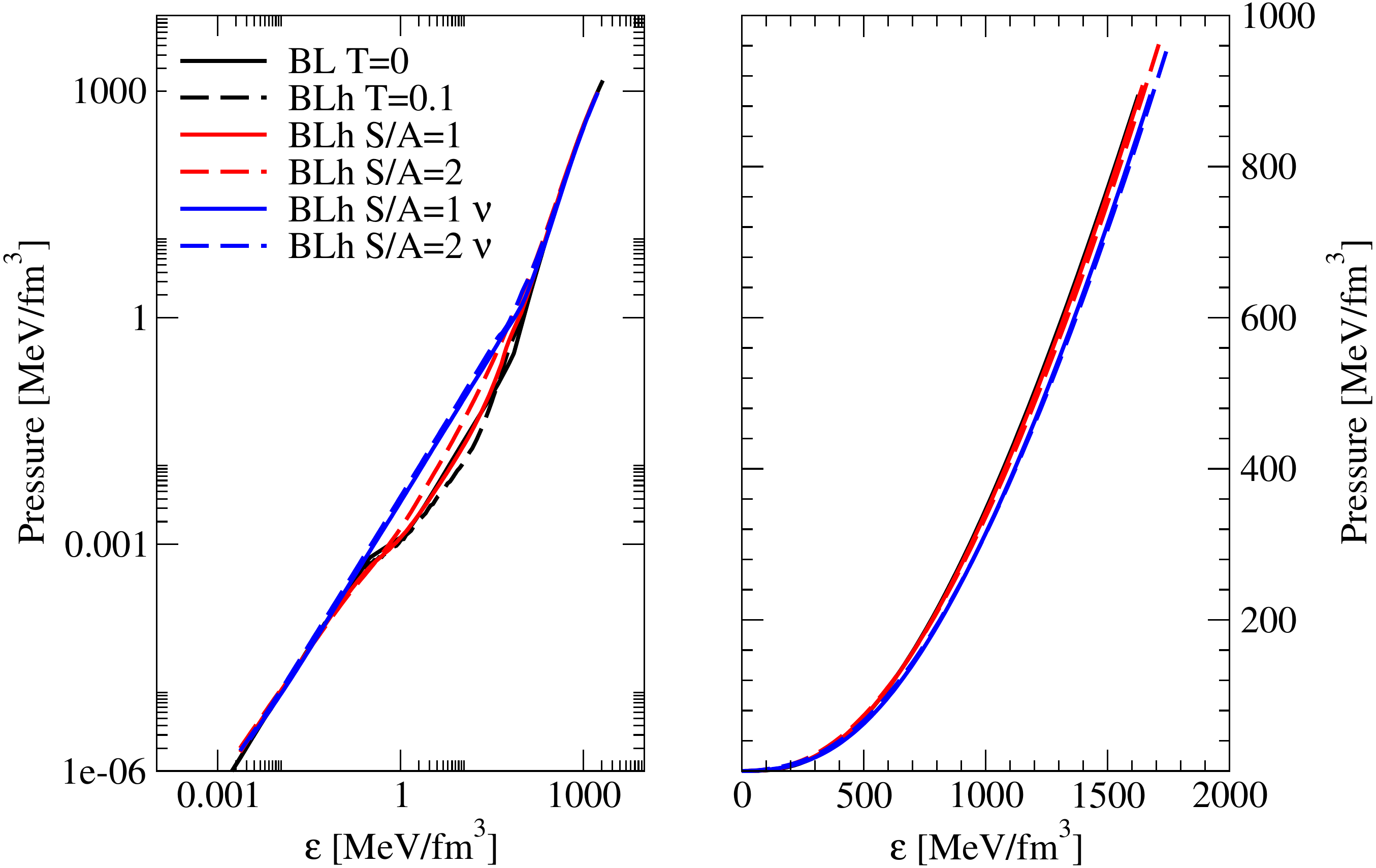} 
\caption{Pressure vs. energy density ($\varepsilon$) relations for various 
$\beta$-stable EOSs derived in different thermodynamical stellar conditions. In the left (right) panel we use a logarithmic  (linear) scale.  
 The BL $T=0$ (black continuous line) taken from \citep{BL} is shown for reference. We note that for the BL EOS for nuclear densities below   
$0.08$ $\rm fm^{-3}$, the Sly4 EOS \citep{sly4_1,sly4_2} has been adopted \citep{endrizzi18}.  
The BLh with $T=0.1\ {\rm MeV}$ is represented by the black dashed line. 
Red and blue curves refer to the following cases: isoentropic EOS with $S/A=1$ and $S/A=2$ without neutrino trapping (red dashed lines)  and isoentropic EOS with $S/A=1$ and $S/A=2$ with neutrino trapping (blue continuous and dashed lines). For the low-density part of the BLh EOSs,  
we employ the TM1 model \citep{TM1}. 
Temperatures (T) in the figure are in {\rm MeV}. (See text for further details).}
\label{eos_beta}
\end{figure}

Figure 7 shows the EOSs (namely the relation between pressure and energy density ($\varepsilon$)) for $\beta$-stable nuclear matter assuming several thermodynamical stellar conditions in logarithmic (left panel) and linear (right panel) scale. 
The two different scales are adopted in order to highlight the low- and high-density parts of each EOS. 
We considered the following cases: $T=0.1$ {\rm MeV}, 
isoentropic with $S/A = 1$ and $S/A = 2$ without neutrino trapping,  
and isoentropic with $S/A = 1$ and $S/A = 2$ with neutrino trapping. 
We also report the $T = 0$ EOS of \cite{BL}  for comparison. We note that  for nuclear densities below   
$0.08$ $\rm fm^{-3}$, the BL EOS is joined with the Sly4 EOS \citep{sly4_1,sly4_2}. 
This combination has been adopted in \citet{endrizzi18} in order to simulate a BNSM of equal-mass neutron stars.
In this work, as explained in the previous section, for $n \leq 0.05\ {\rm fm}^{-3}$, we employ the TM1 model \citep{TM1} (and very similar results were obtained in the case of the SFHo EOS).
We therefore note that for each EOS reported in Fig.\ \ref{eos_beta} we have consistently employed the same thermodynamical conditions assumed for the core as for low-matter density ($n \leq 0.05\ {\rm fm}^{-3}$).  
Additionally, we note that in the case of trapped neutrinos we have enforced the weak equilibrium conditions discussed above in the same density interval. An additional comment about this point is in order. 
Neutrinos rapidly escape from the star in low-density matter. This is usually schematized introducing the 
concept of a neutrino-sphere inside which neutrinos are trapped. However, as discussed in several works \citep[e.g.,][]{stro,fish, endrizzi20}, the exact location of the "radius" of the neutrino-sphere is energy dependent. In the references quoted above this "radius" was found in a range of densities, namely  $10^{-5}-10^{-3}\ {\rm fm}^{-3}$ , for relevant neutrino energies. In the present work, fixing $S/A = 1$ in $\beta$-stable matter, we find that at a nuclear density of around $4.7 \times 10^{-5}\ {\rm fm}^{-3}$ neutrinos disappear. Increasing $S/A,$ this threshold drops slightly; for instance setting $S/A = 2$ we find that the threshold becomes $1.4 \times 10^{-5}\ {\rm fm}^{-3}$. We therefore locate the radius of the neutrino-sphere according to the density at which neutrinos disappear. We consider this criterion to deal with the low-density neutrino component as a natural and reasonable one for our purposes.   

\subsection{Neutron star structure}

\begin{figure}[t]
\centering
\includegraphics[width = 9.cm]{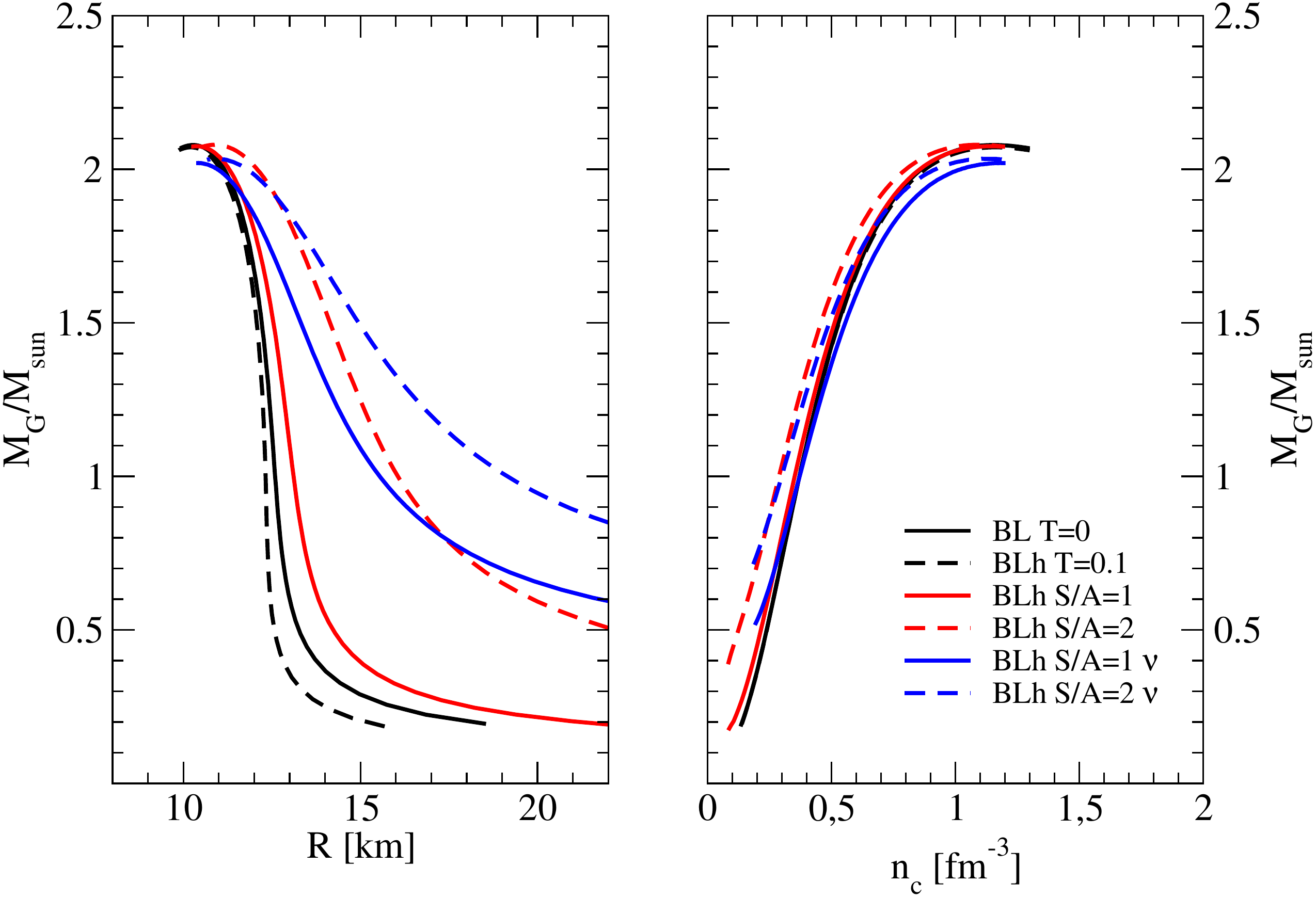} 
\caption{Mass--radius (left panel) and mass--central density ($n_c$) (right panel) for the EOSs reported 
in Fig.\ \ref{eos_beta}. Temperatures (T) in the figure are in {\rm MeV}.}
\label{Mr}
\end{figure}

The structural properties of nonrotating neutron stars can be obtained by integrating the equation for hydrostatic equilibrium in general relativity \citep{Tol34,OV39}. We numerically integrated these equations using the microscopic EOSs of $\beta$-stable nuclear matter described in the previous sections.
For each static neutron star configuration, we additionally compute the total baryon number $N_B$, and we define the baryonic mass (or ``rest mass'') of the neutron star as $M_B = m_u N_B$, where $m_u$ is the baryonic mass unit that we take equal to $m_u = m(^{12}C)/12 = 1.6605 \times 10^{-24} {\rm g}$. The total binding energy of the star is therefore given by $E_{\rm bind} = (M_B - M_G) c^2 $ which represents the total energy liberated during the  birth of the neutron star.  
The results are shown in Fig.\ \ref{Mr}, where we plot the gravitational mass $M_G$ (in unit of solar mass $M_\odot$) as a function of the surface stellar radius $R$ (panel (a)), 
and the gravitational mass as a function of the central density $n_c$ (panel (b)) for the considered EOS models.  
We note that our zero-temperature EOS model (BL) is compatible with the largest current measured neutron star mass, which is 
$M_G = 2.14^{+0.2}_{-0.18} \, M_{\odot}$ \citep{croma19} for the neutron stars in PSR~J0740+6620. 
Various structural properties of the maximum mass configuration for the considered EOS models are listed in Table\ \ref{tab4}.  
We note that thermal effects lead to a monotonic behavior \citep{bomb+1995} of the gravitational maximum mass, both in the case of isoentropic and isothermal EOSs. 
Neutrino trapping instead makes the EOS softer and thus leads to a decrease in the values of both the gravitational and  baryonic maximum mass \citep{bomb96,prak97}. This can be seen when comparing the results in Table\ \ref{tab4} for the 
isoentropic EOS in the neutrino-free and neutrino-trapped (last three rows in Table\ \ref{tab4}) cases. 
This effect was already apparent in the discussion of the composition of $\beta$-stable matter in the neutrino-free and 
neutrino-trapped cases (see Fig. \ref{composition}).   
The electron chemical potential $\mu_e$ in neutrino-trapped matter is indeed increased by the value of the neutrino chemical potential $\mu_{\nu_e}$ (see Eq. \ref{beta1}). In this way, a larger fraction of electrons and thus protons 
(due to charge neutrality) are present at a given nuclear density. Consequently, the EOS becomes softer, being more proton rich. 
However, we note  that the presence of other degrees of freedom in neutrino-trapped matter, like for instance hyperons, may change the previous conclusion \citep{bomb96,prak97}. 
A final remark is that a consistent treatment of neutrinos in the low-density region produces an increase in the minimum neutron-star mass. Looking at Fig.\ \ref{eos_beta} it is clear that for $1 \ {\rm MeV}\ {\rm fm}^{-3} \lesssim \varepsilon \lesssim \ 10^3\ {\rm MeV}\ {\rm fm}^{-3}$ (which corresponds to $10^{-5}\ {\rm fm}^{-3}\lesssim n \lesssim 0.1\ {\rm fm}^{-3}$) the EOSs with trapped neutrinos (blue lines) are stiffer than the ones obtained in the neutrino-less case. This is in agreement with the findings of other works  \citep{burgio10}.   

\begin{table} 
\caption{Maximum mass configuration properties for various thermodynamical stellar conditions considered in this work.}  
\label{tab4}
\small
\centering
\begin{tabular}{ccccc}
\hline
\hline
  EOS model     & $M_G$ ($M_\odot$) & $R$ (km) & $n_c$ (fm$^{-3}$) & $M_B$ ($M_\odot$)  \\
\hline
BL  $T=0$  & 2.080 & 10.26 & 1.156  & 2.456    \\
BLh $T=0.1$& 2.070 & 10.18 & 1.175  & 2.457    \\
BLh $T=5$  & 2.072 & 10.00 & 1.162  & 2.457    \\ 
BLh $T=10$  & 2.073 & 10.05 & 1.158  & 2.457    \\ 
BLh $T=30$  & 2.090 & 10.31 & 1.141  & 2.449    \\ 
BLh $T=50$  & 2.112 & 10.91 & 1.084  & 2.406    \\  
BLh $S/A=1$  & 2.076 & 10.37 & 1.141  & 2.446    \\ 
BLh $S/A=2$  & 2.082 & 10.81 & 1.098  & 2.390    \\
BLh $S/A=3$  & 2.088 & 11.97 & 0.992  & 2.296    \\
BLh $S/A=1$ $\nu$  & 2.021 & 10.39 & 1.190  & 2.323    \\ 
BLh $S/A=2$ $\nu$  & 2.034 & 10.92 & 1.129  & 2.289    \\ 
BLh $S/A=3$ $\nu$  & 2.051 & 12.07 & 1.018  & 2.223    \\ 
\hline
\end{tabular}
\tablefoot{Stellar gravitational maximum mass $M_G$, corresponding radius $R$,  
central baryon number density $n_c$, and baryonic maximum mass $M_B$. 
Stellar masses are given in units of solar mass $M_\odot = 1.989\times 10^{33}$~g.}
\end{table}
%
To test the zero temperature limit of our finite temperature EOS model, we calculated 
an isothermal NS sequence with $T=0.1$ {\rm MeV}. 
As one can see, comparing the results in the first two rows of Table\ \ref{tab4}, 
we recover the properties of the maximum mass configuration predicted by the $T = 0$ BL EOS.   
The differences in the radius (see  Fig.\ \ref{Mr}) for stars with masses 
of less than about  $1\,M_{\odot}$, between the BL EOS and the BLh at $T=0.1$ {\rm MeV,} are due to the different EOS models  
used at low densities ($n_B < 0.05~{\rm fm^{-3}}$). We note in addition that in \citet{BL}, in order to construct  a parametrization of the energy per nucleon for nuclear $E/A$ densities $n\ge 0.08$ {\rm fm}$^{-3}$, only the potential energy part of microscopic calculations was included in the fit. At zero temperature, the kinetic contribution 
to $E/A$ is indeed analytic (Eqs. (6) and (7) in  \citet{BL}).  
In the present work, in which we have extended the microscopic results of \citet{BL} to finite temperature, 
we have instead performed a fit of the full  free energy per nucleon $F/A$ because in this case the kinetic term 
is  also nonanalytic. It is therefore important that the two $\beta$-stable EOSs BL and BLh ($T=0.1$ {\rm MeV}) provide almost indistinguishable results (see right panel of Fig.\ \ref{eos_beta}) in the range $n\ge 0.08$ {\rm fm}$^{-3}$.    

\begin{figure}[t]
\centering
\includegraphics[width = 9.cm]{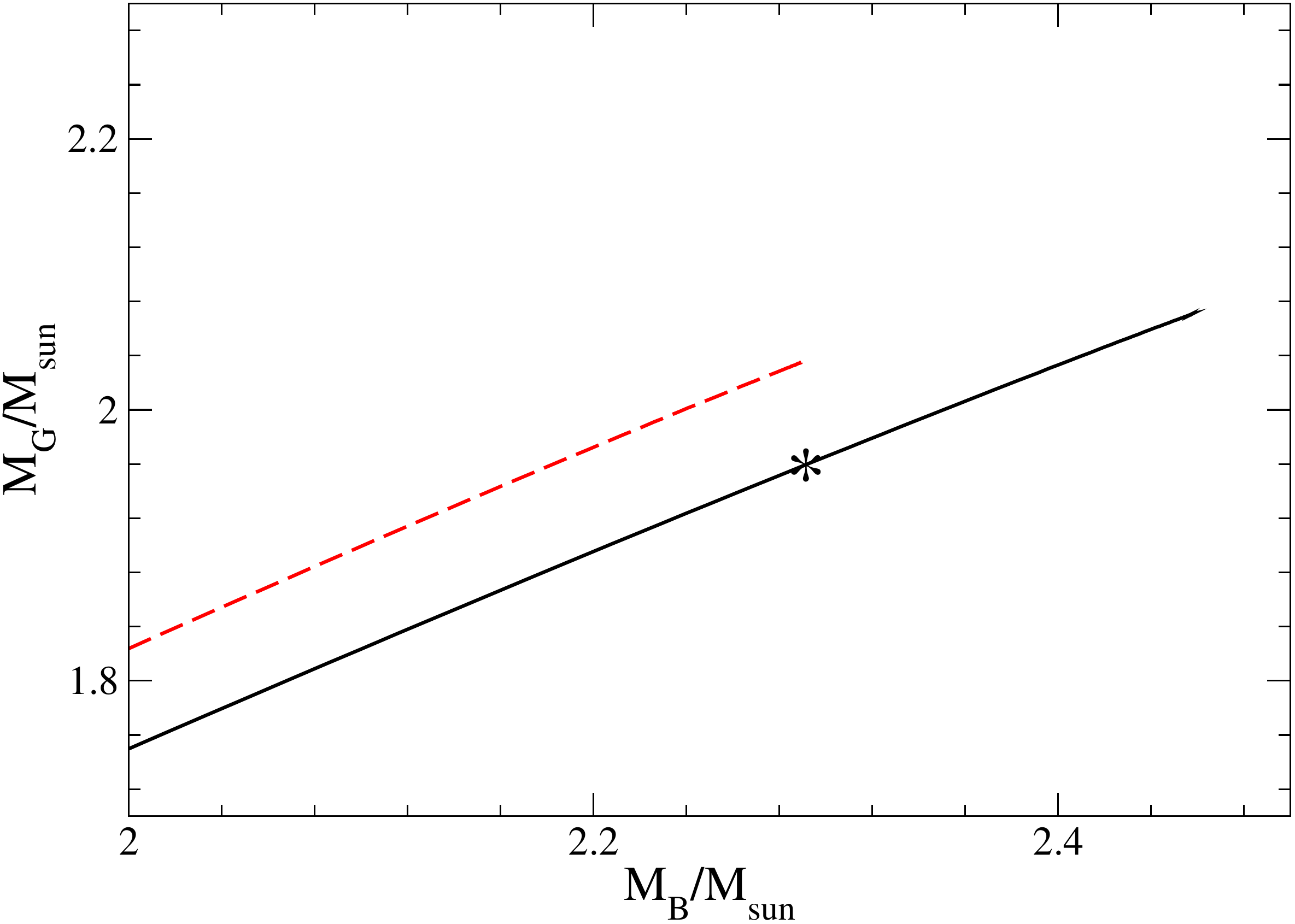} 
\caption{ Gravitational mass $M_G$ as a function of the baryonic mass $M_B$ for our EOS model. 
The red dashed line represents the protoneutron star (PNS) sequence calculated using the isoentropic EOS  
with $S/A = 2$ and trapped neutrinos with $Y_{l_e}=0.33$. 
The black continuous line represents the cold deleptonized NS sequence calculated using the $T = 0.1~{\rm MeV}$ EOS for neutrino-free matter.  
The end point on each curve represents the corresponding maximum mass configuration. 
The values of these quantities for the present BLh EOS model are reported in Table\ \ref{tab4}. 
The asterisk on the black continuous line represents the {\it effective maximum mass} $M_{\rm G\,max}^*$ 
for the cold deleptonized NS sequence (see text for more details). 
}
\label{Mg_Mb}
\end{figure}

\subsection{Early evolution of a newborn  neutron star}

It has been shown \citep{bomb96} that the concept of neutron star maximum mass has to be re-examined
when the formation and early evolution dynamics   
of a newborn PNS are taken into account.
The early evolution of a PNS is driven by thermal and neutrino trapping effects on the EOS. The main features of this process can be schematically investigated considering the following two snapshots of the evolution process:
\begin{itemize}
        \item[(i)] the hot PNS at a time $t \sim 3\,\mathrm{s}$ after core bounce ($t=0$), described by the isoentropic EOS with trapped neutrinos; 
        \item[(ii)] the cold and deleptonized NS at $t \sim 30\,\mathrm{s}$ (neutrino diffusion time), described by the cold and neutrino-free EOS.
\end{itemize}
As most of the matter accretion on the forming NS  happens in the very early stages after birth 
($t < 3\,\mathrm{s}$) \citep{cheavalier89}, the neutron star baryonic mass $M_{\rm B}$ stays almost constant during the evolution between these two configurations. 
The evolution of a PNS can therefore be analyzed in the $M_{\rm G}$--$M_{\rm B}$ plane.
To this purpose we show in Fig.~\ref{Mg_Mb} the gravitational mass as a function of the baryonic mass for the following stellar sequences: (i) PNSs (red dashed line), that is, isoentropic EOS with $S/A = 2$ 
and trapped neutrinos with $Y_{l_e}=0.33$; 
and (ii) cold deleptonized NSs (continuous black line) with $T = 0.1~{\rm MeV}$ and neutrino-free matter. 
The end point on each curve represents the corresponding maximum mass configuration, which we denote as 
$M_{\rm G\,max}^{(i)}$ and $M_{\rm B\,max}^{(i)}$ for the PNS sequence, and with
$M_{\rm G\,max}^{(f)}$ and $M_{\rm B\,max}^{(f)}$ for the final cold deleptonized NS sequence. 
The values of these quantities for the present BLh EOS model are reported in Table\ \ref{tab4}. 
From our results in Fig.\ \ref{Mg_Mb} one can see that a PNS born with $M_{\rm B}^{(i)} \leq M_{\rm B\,max}^{(i)}$ will evolve to a cold deleptonized NS with a gravitational mass  $M_{\rm G}^{(f)}$ and the same baryonic mass $M_{\rm B}^{(i)}$ 
of the initial PNS configuration.       
The binding energy of the star will increase by $\Delta B  = (M_{\rm G}^{(i)} - M_{\rm G}^{(f)})\, c^2$. 
This energy will be released mainly through neutrino emission. In the specific case where $M_{\rm B}^{(i)} = M_{\rm B\,max}^{(i)}$ and for the case considered in Fig.\ \ref{Mg_Mb}, we find $M_{\rm G}^{(f)} = 1.958\,M_\odot$ and $\Delta B = 1.358 \times 10^{53} \mathrm{erg}$.
We now consider a PNS with $M_{\rm B\,max}^{(i)} <  M_{\rm B}^{(i)} \leq M_{\rm B\,max}^{(f)}$\,. 
In this case the PNS cannot be supported by the matter pressure against gravitational collapse 
because its baryonic mass $M_{\rm B}^{(i)}$ is greater than the maximum possible baryonic mass $M_{\rm B\,max}^{(i)}$ for the initial configuration.
Thus, the collapsing stellar core will collapse to a BH after reaching supranuclear densities.  
In the classical analysis, {\it à la}  Oppenheimer--Volkoff, where the dynamical evolution of the PNS is not taken into account, stars in this baryonic mass range are considered to have a stable equilibrium configuration in the cold deleptonized stellar sequence. 
Thus, the gravitational mass $M_{\rm G\,max}^* \equiv M_{\rm G}^{(f)}(M_{\rm B\,max}^{(i)})$ of the star, corresponding to the  evolution of the maximum mass PNS configuration, plays the role of an {\it effective maximum mass} for the cold deleptonized NS sequence \citep{bomb96}. 
We stress that this is a feasible mechanism to produce low-mass black holes ($M_{\rm BH} \sim 2\,M_{\odot}$) and could have far-reaching consequences for interpreting the gravitational wave event GW190814 \citep{GW190814} 
as a BH--BH merger.
Finally, when $M_{\rm B}^{(i)} >  M_{\rm B\,max}^{(f)}$ the stellar core will collapse to a BH. 

Consider now a NS in a binary stellar system in which the companion star is a normal star. If  a common envelope is formed during the evolution of the binary system, the NS can accrete  matter from its companion star with a certain accretion rate $\dot{M}_{\rm B}$. After a sufficiently long time the NS is able to increase its baryonic mass above the value $M_{\rm B\,max}^{(i)}$ and begins to populate the portion of the final NS sequence with $M_{\rm B\,max}^{(i)} <  M_{\rm B}  \leq M_{\rm B\,max}^{(f)}$\,, eventually reaching and then overcoming the Oppenheimer--Volkoff maximum mass configuration   $M_{\rm B\,max}^{(f)}$\,. 
Thus, as pointed out for the first time in  \citep{bomb96}, in the baryonic mass range 
$[M_{\rm B\,max}^{(i)}\;, M_{\rm B\,max}^{(f)}]$ one can have both NSs and BHs. For the case depicted in Fig.\ \ref{Mg_Mb}, this range is [2.289\,,\,  2.457]\,$M_\odot$\,.

\subsection{Application to spherically symmetric core-collapse supernovae }
\begin{figure*}[th!]
\centering 
\includegraphics[height = 5.7cm]{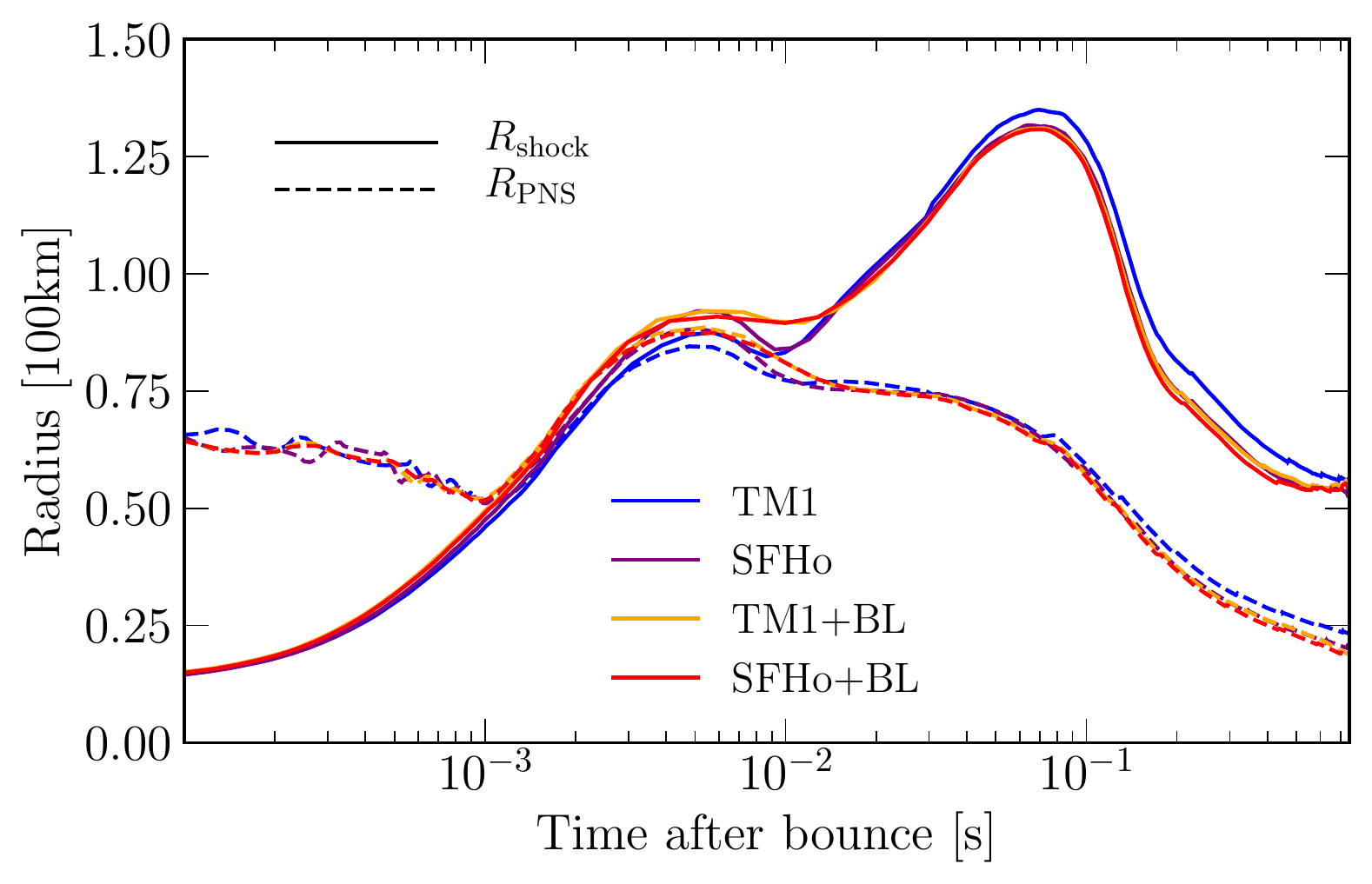}
\includegraphics[height = 5.7cm]{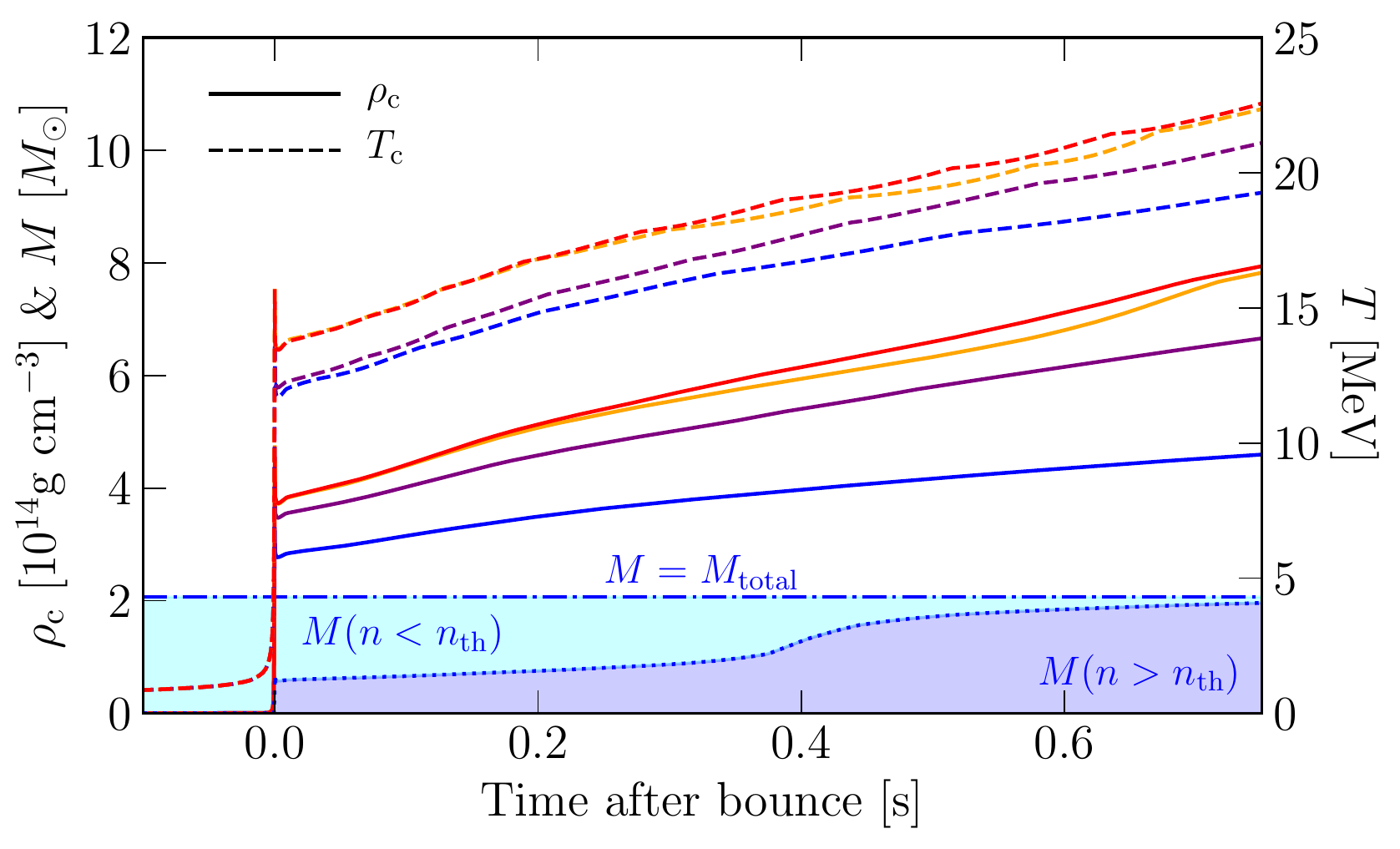}
\caption{Left: Temporal evolution of the shock radius (solid) and of the PNS radius (dashed) for each of the four spherically symmetric CCSN models. Models differ by the nuclear EOSs, including two versions of the BLh EOS and two RMF EOSs, namely HS(TM1) and SFHo. The two BLh EOSs use the HS(TM1) and the SFHo EOS for the low-density regime ($n < n_{\rm th}$). We note the logarithmic scale on the time axis. Right: Same as on the left panel, but for the central density (solid) and temperature (dashed). The cyan (blue) shaded area represents matter below (above) the threshold density $n_{\rm th}$ (used to join the BLh and the HS(TM1) EOS) for the BLh+TM1 model.}
\label{fig:CCSN_radii_central_properties}
\end{figure*}

\begin{figure*}[th!]
\centering
\includegraphics[height = 5.7cm]{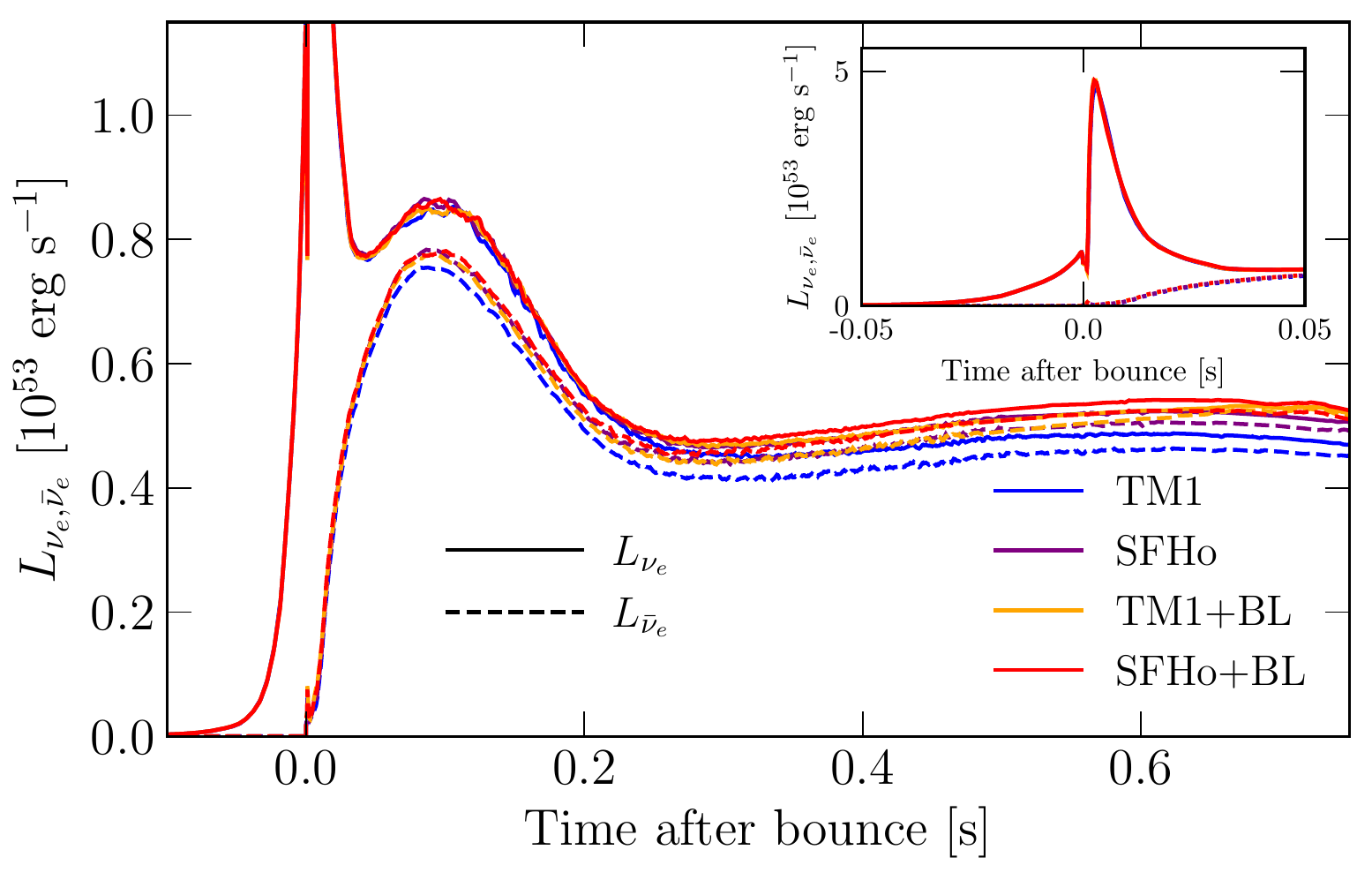}
\includegraphics[height = 5.7cm]{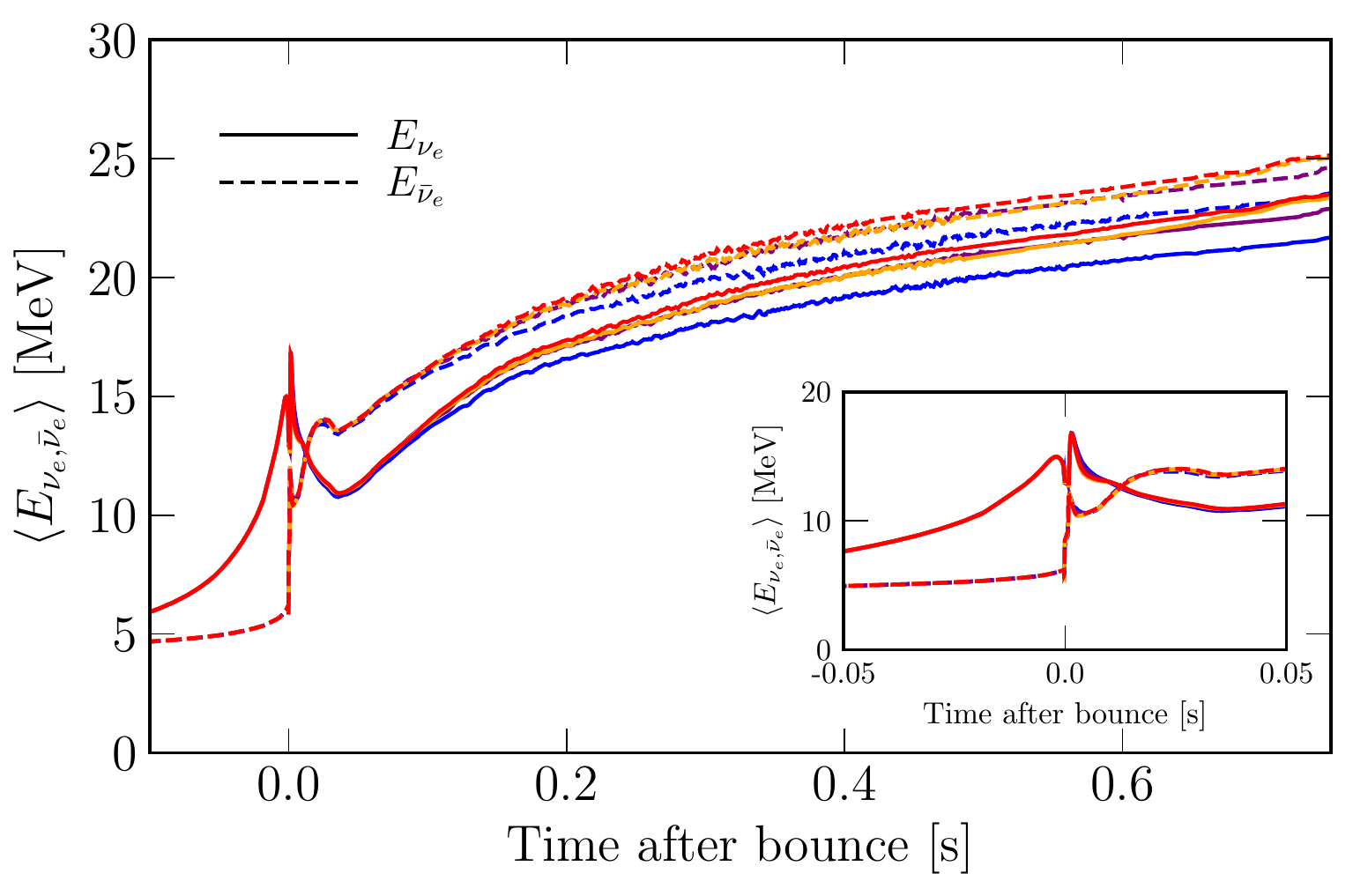}
\caption{Left: Same as in Figure~\ref{fig:CCSN_radii_central_properties}, but for the electron neutrino (solid) and antineutrino (dashed) luminosities measured at the edge of the computational domain.
Right: Same as in the left panel, but for the electron neutrino (solid) and antineutrino (dashed) mean energies.}
\label{fig:CCSN_neutrino_quantities}
\end{figure*}

%
To test our finite temperature EOS in an astrophysical dynamical scenario, we set up simulations of a CCSN in spherical symmetry. For recent reviews of the physics and the status of CCSN modeling we refer to \citet{burrows_review_2013},  \citet{janka_review_2016}. In the following, we summarize only the most significant aspects necessary to understand results from our simulations.

A CCSN marks the end of the life of a massive star, with a zero-age main sequence mass above $\sim 8 M_{\odot}$, up to the pair instability limit. 
During the so-called silicon-burning phase, in the centre of the star silicon is converted into iron group elements in NSE conditions. The forming iron core is mainly sustained by the pressure of degenerate electrons. However, once the Chandrasekhar mass limit has been reached, the gravitational collapse becomes unavoidable and the core collapses under its own gravity. 
At the onset of the collapse, matter in the center has already developed a significant deviation from isospin symmetry ($Y_e \lesssim 0.48$) and typical densities and temperatures are $\rho_c \sim$ $(3-8)$ $\times$ $10^9{\rm g~cm^{-3}} $ and $T_c \gtrsim 0.5~{\rm MeV}$ , respectively, with an intrinsic variability due to the progenitor initial mass, metallicity, and possibly binarity.
The core collapses on the dynamical timescale of
$\tau \sim \left( G \rho_c \right)^{-1/2} \sim 100~{\rm ms} $ and the collapse is further accelerated by the core deleptonization, through electron captures on nuclei and protons, and the subsequent escape of the produced neutrinos. While neutrinos initially escape freely, only when $\rho_c \sim 10^{12}{\rm g~cm^{-3}}$ does the typical neutrino mean free path decrease below the size of the inner core and neutrinos start to be trapped.
The collapse proceeds until $\rho_c \sim \rho_0$. The subsequent EOS stiffening halts the collapse and produces an outgoing shock wave. The moment when the maximum density is reached is called core-bounce.
At core bounce, the electron fraction at the  center of the core is $Y_e \lesssim 0.3$, while the fraction of trapped electron neutrino is $Y_{\nu_e} \lesssim 0.1$.
Temperatures increase up to a few tens of MeV and the shock photodissociates the iron core into free neutrons and protons. Neutrinos of all flavors are copiously produced and emitted with typical neutrino luminosities of the order of a few times $10^{52}{\rm erg~ s^{-1}}$.
After a few tens of milliseconds,  the shock loses its energy because of photodissociation and neutrino
emission, and then stalls and converts into an accretion shock, inside which a PNS forms. 
The neutrino diffusion from the optically thick PNS, and the persistent deleptonization and compression of the accreting matter guarantee high neutrino luminosities over a timescale of one second. 
The re-absorption of electron flavor (anti)neutrinos mainly emitted at the surface of the PNS produces a region of net energy deposition just behind the shock (the so-called gain region). This energy deposition is thought to be the cause of the shock revival that eventually explodes the star.
Detailed models in spherical symmetries including all the relevant physics fail to revive the shock through neutrino heating.
Multi-dimensional simulations proved that convection and hydrodynamical instabilities play a crucial role in helping the explosion to occur.

Our hydrodynamic simulations are performed with the Lagrangian general-relativistic, adaptive-grid code \textit{Agile} \citep{agile_paper}.
We employ the isotropic  diffusion  source approximation (IDSA) for the transport of electron (anti)neutrinos \citep{idsa_paper}.
The neutrino energy is discretized using $20$ geometrically spaced energy bins in the range $\left[ 3, 300 \right]$ MeV. The neutrino reactions included in the IDSA are electron captures on nuclei, electron and positron captures on free nucleons, and scattering off free nucleons and nuclei \citep{Bruenn1985,Mezzacappa1993}. These represent the  minimal  set  of  the  most relevant weak processes during the collapse and in the early post-bounce phase.
The emission of $\mu$ and $\tau$ (anti)neutrinos in the post-bounce phase is handled by a gray leakage scheme \citep{Cabezon2018}. This scheme has been gauged on detailed Boltzmann transport models, and includes pair annihilation, nucleon--nucleon bremsstrahlung, and scattering off nucleons.
With this setup, we simulate the collapse, bounce, and post-bounce phases (roughly the first 700ms after bounce) for a 20$M_{\odot}$ ZAMS, solar metallicity progenitor star \citep{WoosleyHeger2007}. We have chosen this progenitor because it has been extensively explored in several recent comparison papers in spherical symmetry and in multi-dimensions \citep{oconnor18,Pan2019,Schneider2019}.
For this specific progenitor, at the onset of the collapse, $\rho_c \approx 3.3 \times 10^{9}{\rm g~cm^{-3}}$, $T_c \approx 0.67 {\rm MeV,}$ and $Y_{e,c} \approx 0.44 $. In all our models we use 103 radial zones which includes the innermost $\sim 2.1 M_{\odot}$. This corresponds to an initial radius of $R \approx 5 \times 10^8{\rm cm}$.
We finally stress that the code that we are using is identical to the publicly available version of AGILE-IDSA\footnote{https://astro.physik.unibas.ch/people/matthias-liebendoerfer/download.html}. 

To close the set of hydrodynamics equations and to compute the necessary neutrino emissivities, absorptivities, and opacities, we couple a finite-temperature, composition-dependent nuclear EOS to the CCSN model. We consider four different EOSs: two versions of the complete BLh EOS and two RMF EOSs, namely the SFHo and HD(TM1) EOS. The two versions of the BLh EOS differ in the EOS used for the low-density part ($n_B < 0.05~{\rm fm^{-3}}$) joined to the high-density BLh EOS according to the procedure described in 
Section~\ref{EOS hadronic}. All these EOSs include electron, positrons, and photons as outlined at the beginning of this section. 
In particular, we used precisely the same SFHo and HS(TM1) EOSs. Thus, our choice allows us to directly address the impact of the transition to the low-density EOS, as well as the difference between our EOS and other EOSs commonly used in the literature. 
Trapped neutrinos are consistently modeled by the IDSA, while muons are neglected.
We stress that the use of a CCSN model to test the transition is challenging, because all matter inside the domain is initially at low densities and most of it has dynamically transited to high densities by the end of the simulation. 

All the different models run without problems through the different CCSN phases (collapse, bounce, after-bounce). Core bounce happens $\sim$378.6(384.6) ms after the beginning of the simulation for the SFHo and TM1 (for the SFHo+BLh and TM1+BLh) models.
In Figures~\ref{fig:CCSN_radii_central_properties} and \ref{fig:CCSN_neutrino_quantities} we present the outcome of our set of simulations.
In particular, in the left panel of Figure \ref{fig:CCSN_radii_central_properties} the temporal evolution of the shock radius and the PNS radius are depicted. The latter is defined as the radius where the rest mass density is equal to $10^{11}{\rm g~cm^{-3}}$. In the right panel of the same figure, we present the evolution of the central density and temperature. In Figure \ref{fig:CCSN_neutrino_quantities} the neutrino luminosities and mean energies are reported from the latest phase of the collapse up to the end of the simulation.

The CCSN models employing the BLh high-density EOS show all the features expected from a spherically symmetric CCSN. The PNS radius, shock radius, neutrino luminosities, and mean energies follow the expected trends, with all the models being  close to one another throughout the simulated time. When the central density reaches values above nuclear saturation density, the core bounces, and an outgoing shock wave forms. The shock expands up to $\sim$135km and then recedes and converts into an accretion shock.  A hot mantle is built below
the shock, and  the forming PNS contracts inside it due to the emission of neutrinos. 

Matter in the center compresses, steadily increasing the central density. Due to compression, the core temperature also increases. As the neutrino diffusion time from the PNS is of the order of seconds (much larger than the simulated time), the lepton fraction, the electron fraction, and the entropy in the center remain almost constant during the simulation ($Y_l \approx 0.32-0.34$ , $Y_e \approx 0.28-0.30$ and $S/A \approx 1.25-1.35 $). Matter in the center is highly degenerate and the increase in temperature is relatively small ($T_c \lesssim 20-30~{\rm MeV}$, even at the end of the simulation) compared to the corresponding Fermi temperature.
Among the different EOSs, HS(TM1) is the stiffest and its central densities are significantly lower than for the other EOSs. The largest densities are reached by the BLh-based EOSs, with very minor differences between its two different implementations. The BLh models  also present the largest central temperature. For this quantity, the relative difference between BLh+SFHo and BLh+TM1 varies slightly with time, with the latter being very close to the former close to bounce and towards the end of the simulation. Due to the lower densities and temperatures reached by the TM1 model, its shock and PNS radii are systematically larger than for other EOS models, while the neutrino luminosities and mean energies are systematically smaller. Apart from this expected difference, we note how the SFHo+BLh and TM1+BLh stay close to each other during the entire simulation, including during the $\nu_e$ burst peak and the corresponding rapid shock expansion phase, between 1 and 10 ms after bounce.
Finally, in the lower part of the right panel of Figure~\ref{fig:CCSN_radii_central_properties}, by the end of the simulation, for models using the BLh-based EOS, most of the matter in the computational domain is described by the high-density part of the EOS. 

\section{Summary}

In this work, we extend the zero-temperature microscopic EOS of dense asymmetric and $\beta$-stable nuclear matter of \citet{BL} to finite temperature.   
For consistency, we used the same many-body method as well as the same two- and three-body nuclear forces of\ \cite{BL}. 
For the low-ensity part of our EOS (namely below $0.05\ {\rm fm^{-3}}$), we connected our microscopic EOS with a EOS derived in the RMF approach \citep{hempel10}. In order to test different choices, we used two publicly available EOS tables obtained from two different parametrizations of the  nuclear RMF Lagrangian: the TM1 \citep{TM1} and SFHo \citep{SFHO} ones. 

We considered three applications of this new EOS.   
The first one concerned the calculation of the isoentropic $\beta$-stable EOS considering both the case of $\nu$-free and $\nu$-trapped matter. We note that the effects of neutrino trapping has been analyzed in the whole range of density of interest for hot neutron stars, therefore including the low-density EOS part. In our second application, we used the finite temperature $\beta$-stable EOSs to describe the structure of nonrotating protoneutron stars. We discuss the thermal effects on the maximum and minimum neutron star masses obtained according to the various physical scenarios analyzed in this paper. 
In our last application, we employed our new EOS to perform a dynamical simulation of spherically symmetric CCSNe. This last application  turned out to be particularly important because it is very sensitive to the density range where we match microscopic calculations with the RMF results. Our findings are in agreement with those present in the literature on this kind of simulation, supporting the plausibility of the EOS construction proposed in this paper.  

We note in addition that the zero-temperature neutron star maximum mass predicted by our new EOS is consistent with present measured neutron star masses and particularly with the mass $M = 2.14^{+0.20}_{-0.18} \, M_{\odot}$ of the neutron stars in PSR~J0740+6620.  

The consistent thermodynamical approach adopted in this work is relevant for multidimensional CCSNe and BNSM simulations \citep{shen98,hempel10,SFHO,oertel2017,togashi17}. 
In the near future we  indeed plan to apply this new EOS to BNSMs as well as to 3D CCSN simulations. 

We also plan to develop a more consistent treatment of the low-density part of our EOS namely below $0.05 {\rm fm^{-3}}$. A possibility may consist in finding a skyrme parametrization able to fit microscopic calculations around saturation density and to join the high-density part of the EOS based on fully microscopic calculations with the latter.   

We finally note that due to the very high central densities expected in the core of NSs, various ``exotic'' constituents like hyperons \citep{hyp1,hyp2,hyp3,hyp4} or a deconfined quark phase \citep{gle96,bombaci2008,log2012,bl2013,bombaci2016} cannot be excluded.  Thermal effects may favor the presence of these additional degrees of freedom in neutron star matter. Work in this direction is ongoing.  Specifically, we are currently trying to extend our present calculations to the case of finite-temperature hyperonic matter, adopting nucleon-hyperon \citep{ny_chiral} and hyperon-hyperon \citep{yy_chiral} interactions derived in the framework of ChPT, and to include hyperonic three-body forces like those in the zero-temperature case discussed by  \citet{logoteta_nny}.

\section*{Acknowledgement}  
This work has been partially supported by ``PHAROS'', COST Action MP1304. AP acknowledges the usage of computer resources under a CINECA-INFN (allocation INF20\_teongrav).




\begin{thebibliography}{} 


\bibitem[Abbott et al. (2017)]{gw5}
Abbott, B. P. et al. (LIGO Scientific Collaboration and Virgo Collaboration) 2017, Phys. Rev. Lett., 119, 161101

\bibitem[Abbott et al. (2018)]{gw170817_eos} Abbott, B. P. et al. (LIGO Scientific Collaboration and Virgo Collaboration) 2018, Phys. Rev. Lett., 121, 161101

\bibitem[Abbott et al. (2020a)]{gw6}
Abbott, B. P. et al. (LIGO Scientific Collaboration and Virgo Collaboration) 2020, ApJ Lett., 892, L3

\bibitem[Abbott et al. (2020b)]{GW190814}
Abbott, B. P. et al. (LIGO Scientific Collaboration and Virgo Collaboration) 2020, ApJ Lett., 896, L44

\bibitem[Akmal et al. (1998)]{apr98}   
Akmal, A., Pandharipande V. R., \& Ravenhall D. G., \ 1998, Phys. Rev. C, 58, 1804  

\bibitem[Antoniadis et al. (2013)]{anto2013} 
Antoniadis, J. et al. \ 2013 Science, 340, 1233232  

\bibitem[Baiotti \& Rezzolla (2017)]{baiotti17}
Baiotti, L., \& Rezzolla, L. \ 2017, Rep. Progr. in Phys., 80, 9  

\bibitem[Baldo et al. (1997)]{bbb97}    
Baldo M., Bombaci I., \& Burgio G. F. \ 1997, A\&A , 328, 274 

\bibitem[Baldo et al. (1991)]{baldo+91}  
Baldo M., Bombaci I., Ferreira L. S.,  Giansiracusa G., \&  Lombardo U. \ 1991, Phys. Rev. C, 43, 2605


\bibitem[Baldo et al. (1990)]{baldo+90}  
Baldo, M.,  Bombaci I.,  Giansiracusa G.,  Lombardo U.,  Mahaux C., \&  Sartor R. \ 1990   
                      Phys. Rev. C, 41, 1748 

\bibitem[Baldo \& Ferreira (1999)]{BF99} 
Baldo, M., \& Ferreira, L.S., Phys. Rev. C, \ 1999, 59, 682 

\bibitem[Baldo et al. (2000)]{baldo00}  
Baldo M.,  Giansiracusa G.,  Lombardo U., \& Song H. Q. \ 2000, Phys. Lett. B, 473, 1 

\bibitem[Bauswein \& Janka (2012)]{BJ2012} 
Bauswein, A., \& Janka, H.-T. \ 2012, Phys. Rev. Lett., 108, 011101

\bibitem[Bauswein et al. (2010)]{BJO2010} 
Bauswein, A., Janka, H.-T. \& Oechslin, R. \ 2010, Phys. Rev. D, 82, 084043

\bibitem[Bauswein et al.(2019)]{Bauswein2019} Bauswein, A., Bastian, N.-U.~F., Blaschke, D.~B., et al.\ 2019, \prl, 122, 061102

\bibitem[Bernuzzi et al. (2015)]{Bernuzzi2015} 
Bernuzzi, S., Dietrich, T, \& Nagar, A. \ 2015, Phys. Rev. Lett., 115, 091101 

\bibitem[Bernuzzi et al.(2016)]{Bernuzzi2016} Bernuzzi, S., Radice, D., Ott, C.~D., et al.\ 2016, \prd, 94, 024023

\bibitem[Bernuzzi et al. (2020)]{Bernuzzi20} 
Bernuzzi, S., Breschi, M., Daszuta, B., Endrizzi, A., Logoteta, D., Nedora, V., Perego, A., Radice, D., 
Schianchi, F., Zappa, F., Bombaci, I., \&  Ortiz, N. \ 2020, \mnras, 497, 1488

\bibitem[Bethe (1965)]{bethe65}  Bethe, H. A. \ 1965, Phys. Rev., 138, 804B 

\bibitem[Bethe et al. (1963)]{BBP63} 
Bethe, H. A., Brandow, B. H., \& Petschek, A. G. \ 1963, Phys. Rev., 129, 225 

\bibitem[Binder et al. (2016)]{binder16}  
Binder, S. et al. (LENPIC Collaboration) \ 2016, Phys. Rev. C, 93, 044002 

\bibitem[Bloch \& de Dominicis (1958)]{BD58} 
Bloch, C., \& De Dominicis, C. \ 1958, Nucl. Phys. A, 7, 459   

\bibitem[Bloch \& de Dominicis (1959a)]{BD59} 
Bloch, C., \& De Dominicis, C. \ 1959, Nucl. Phys. A, 10, 181   

\bibitem[Bloch \& de Dominicis (1959b)]{BD59_1} 
Bloch, C., \& De Dominicis, C. \ 1959, Nucl. Phys. A, 10, 509   

\bibitem[Bombaci (1996)]{bomb96} Bombaci, I. \ 1996, A\&A, 305, 817 

\bibitem[Bombaci et al. (1995)]{bomb+1995}  Bombaci, I., Prakash, M., Prakash, M., Ellis, P.J., Lattimer, J.M., 
\& Brown, G.E. \ 1995, Nucl. Phys. A 583, 623  

\bibitem[Bombaci et al. (2008)]{bombaci2008} 
Bombaci, I., Panda, P. K., Provid{\^e}ncia, C., \& Vida\~na, I. \ 2008, Phys. Rev. D, 77, 083002

\bibitem[Bombaci \& Logoteta (2013)]{bl2013} 
Bombaci, I., \&  Logoteta, D. 2013, MNRAS Lett. 433, L79

\bibitem[Bombaci et al. (2016)]{bombaci2016}  
Bombaci, I., Logoteta, D., Vida\~na, I., \& Provid{\^e}ncia, C. \ 2016, Eur. Phys. J. A, 52, 58

\bibitem[Bombaci \& Logoteta (2018)]{BL} 
Bombaci, I., \&  Logoteta, D. \ 2018, A\&A, 609, A128 

\bibitem[Bombaci \& Lombardo (1991)]{bl91} Bombaci, I., \& Lombardo, U. \ 1991, Phys. Rev. C, 44, 1892 

\bibitem[Bombaci et al. (1993)]{bkl93} Bombaci, I., Kuo, T. S., \& Lombardo, U. \ 1993, Phys. Lett., B 311, 9 

\bibitem[Bombaci et al. (1994)]{bl94} Bombaci, I., Kuo, T. S., \& Lombardo, U. \ 1994, Phys. Rep., 242, 165 

\bibitem[Bombaci et al. (2006)]{bomb+06} 
Bombaci, I., Polls, A., Ramos, A., Rios, A., \& Vida\~na, I. \ 2006, Phys. Lett. B, 632, 638  

\bibitem[Bruenn(1985)]{Bruenn1985} Bruenn, S.~W.\ 1985, \apjs, 58, 771

\bibitem[Burgio \& Schulze (2010)]{burgio10}
Burgio, G. F., \& Schulze, H.-J. \ 2010, A\&A 518, A17 

\bibitem[Burgio et al. (2011)]{burgio2011}
Burgio, G. F., Schulze, H.-J., \& Li, A. \ 2011, Phys. Rev. C, 83, 025804

\bibitem[Burrows(2013)]{burrows_review_2013} Burrows, A.\ 2013, Reviews of Modern Physics, 85, 245

\bibitem[Burrows \& Lattimer (1986)]{BL86} 
Burrows, A. \& Lattimer, J. M. \ 1986, ApJ, 307, 178

\bibitem[Cabez{\'o}n et al.(2018)]{Cabezon2018} Cabez{\'o}n, R.~M., Pan, K.-C., Liebend{\"o}rfer, M., et al.\ 2018, \aap, 619, A118

\bibitem[Carbone et al. (2018)]{CPR_2018} Carbone, A., Polls, A., \& Rios, A. \ 2018, Phys. Rev. C, 98, 025804 

\bibitem[Carbone \& Schwenk (2019)]{carbone-schwenk_2019} Carbone, A., \&  Schwenk,  A. \ 2019, Phys. Rev. C, 100, 025805 

\bibitem[Chamel et al. (2011)]{chamel11}  
Chamel, N., Fantina, A. F., Paearson, J. M., \& Goriely, S. \ 2011, Phys. Rev. C, 84, 062802(R) 

\bibitem[Chatterjee \& Vida\~na (2016)]{hyp3} 
Chatterjee, D. \& Vida\~na, I. \ 2016, Eur. Phys. J. A, 52, 29 

\bibitem[Chevalier (1989)]{cheavalier89} Chevalier, R.A. \ 1989, ApJ, 346, 847

\bibitem[Coester et al. (1970)]{coester70} 
Coester F.,  Cohen S.,  Day B., \&  Vincent C. M. \ 1970, Phys. Rev. C, 1, 769  

\bibitem[Cromartie et al. (2019)]{croma19}
Cromartie, H. T. et al. \ 2019, Nature Astronomy 10.1038 

\bibitem[Day (1967)]{bbg1} Day, B. D. \ 1967, Rev. Mod. Phys.,  39, 719 

\bibitem[Day (1981)]{day81} Day B. D. \ 1981, Phys. Rev. Lett.,  47, 226 

\bibitem[Demorest et al. (2010)]{demo2010} Demorest P., Pennucci T., Ransom S., Roberts M., Hessels J. 
        \ 2010, Nature, 467, 1081 

\bibitem[Douchin \& Haensel (2001)]{sly4_2} 
 Douchin, F., \& Haensel, P. \ 2001, A\&A 380,  151  
 
\bibitem[Drischler et al. (2016)]{drisch_2016}  Drischler, C., Hebeler, K., \& Schwenk, A. \ 2016, Phys. Rev. C, 93, 054314 

\bibitem[Drischler et al. (2019)]{drisch19} 
 Drischler, C., Hebeler, K., \& Schwenk, A. \ 2019, Phys. Rev. Lett., 122, 042501 

\bibitem[Endrizzi et al. (2018)]{endrizzi18} 
Endrizzi, A., Logoteta, D., Giacomazzo, B., Bombaci, I.,  Kastaun, W., \&  Ciolfi, R. \ 2018, Phys. Rev. D, 98, 043015

\bibitem[Endrizzi et al. (2020)]{endrizzi20} Endrizzi, A., Perego, A.,  Fabbri, F. M., Branca, L., Radice, D., Bernuzzi, S., Giacomazzo, B., Pederiva, F. \& Lovato, A. \ 2020, Eur. Phys. J. A, 56, 15

\bibitem[Fattoyev et al. (2010)]{IUFSU} 
Fattoyev, F. J., Horowitz, C. J., Piekarewicz, J., \& Shen, G. \ 2010, Phys. Rev. C, 82, 055803 

\bibitem[Fiorilla et al. (2012)]{fiorilla12} 
Fiorilla, S., Kaiser, N. \& Weise, W. \ 2012, Nucl. Phys. A, 880, 65   

\bibitem[Fisher et al. (2009)]{fish}
Fischer, T., Whitehouse, S. C., Mezzacappa, A., Thielemann, F.-K., \&
Liebend\"orfer, M.\ 2009, A\&A, 499, 1

\bibitem[Fischer et al.(2010)]{Fischer2010} Fischer, T., Whitehouse, S.~C., Mezzacappa, A., et al.\ 2010, A\&A, 517, A80

\bibitem[Fischer et al.(2014)]{Fischer_2014} Fischer, T., Hempel, M., Sagert, I., et al.\ 2014, European Physical Journal A, 50, 46

\bibitem[Friedman \& Pandharipande (1981)]{FP81}     
Friedman, B., \&  Pandharipande, V. R. \ 1981, Nucl. Phys. A, 361, 502 

\bibitem[Frick \& M\"uther (2003)]{FM03}
Frick, T., \& M\"uther, H. \ 2003, Phys. Rev. C, 68, 034310 

\bibitem[Frick et al. (2005)]{FM05}
Frick, T., Muther, H., Rios, A., Polls, A., \& Ramos, A. Phys. Rev. C, 71, 014313 

\bibitem[Glendenning (1985)]{hyp1} Glendenning, N.K.\ 1985, ApJ., 293, 470

\bibitem[Glendenning (1996)]{gle96}  Glendenning,  N.K., 1996,  Compact Stars: Nuclear Physics, 
                               Particle Physics, and General Relativity, Springer Verlag 

\bibitem[Grang\`e et al. (1987)]{gra87} 
Grang\'e, P., Cugnon, J., \& Lejeune, A. \ 1987, Nucl. Phys. A, 473, 365    

\bibitem[Haensel \& Pichon (1994)]{sly4_1} 
Haensel, P., \& Pichon, B. \ 1994,  A\&A 283,  313

\bibitem[Haidenbauer et al. (2017)]{hyp4} 
Haidenbauer, J., Mei{\ss}ner, U.-G., Kaiser, N., \& Weise, W. \ 2017, Eur. Phys. J. A, 53, 121 

\bibitem[Haidenbauer et al. (2016)]{yy_chiral} 
Haidenbauer, J., Mei{\ss}ner, U.-G. \& Petschauer, S. \ 2016, Nucl. Phys. A, 954, 273   

\bibitem[Haidenbauer et al. (2013)]{ny_chiral} 
Haidenbauer, J., Petschauer, S., Kaiser, N., Mei{\ss}ner, U.-G., Nogga, A., \& Weise, W. 
\ 2013, Nucl. Phys. A, 915, 24  


\bibitem[Hammer et al. (2013)]{hammer13}   
Hammer, H. W., Nogga, A., \&  Schenk A. \ 2013, Rev. Mod. Phys., 85, 197  

\bibitem[Hanauske \& Bovard(2018)]{Hanauske_2018} Hanauske, M., \& Bovard, L.\ 2018, Journal of Astrophysics and Astronomy, 39, 45

\bibitem[Hempel \& Schaffner-Bielich (2010)]{hempel10}
Hempel, M., \& Schaffner-Bielich, J. \ 2010, Nucl. Phys. A, 837, 210

\bibitem[Holt et al. (2010)]{holt} 
Holt, J. W., Kaiser, N., \&  Weise, W. \ 2010, Phys. Rev. C, 81, 024002    

\bibitem[Hotokezaka et al.(2013)]{Hotokezaka2013} Hotokezaka, K., Kiuchi, K., Kyutoku, K., et al.\ 2013, \prd, 88, 044026

\bibitem[H{\"u}depohl et al.(2010)]{Hudepohl2010} H{\"u}depohl, L., M{\"u}ller, B., Janka, H.-T., et al.\ 2010, \prl, 104, 251101

\bibitem[H\"ufner \& Mahaux (1972)]{HM72} H\"ufner, J., \& Mahaux, C. \ 1972, Ann. Phys. (N.Y.), 73, 525 

\bibitem[Hugenholtz \& Van Hove (1958)]{HVH} Hugenholtz, N. M., \& Van Hove, L. \ 1958, Physica (Amsterdam), 24, 363 

\bibitem[Janka et al.(2016)]{janka_review_2016} Janka, H.-T., Melson, T., \& Summa, A.\ 2016, Annual Review of Nuclear and Particle Science, 66, 341

\bibitem[Jeukenne et al. (1976)]{jeuk+67}  
Jeukenne, J. P., Lejeunne, A., \& Mahaux, C. \ 1976, Phys. Rep., 25, 83  

\bibitem[Kalantar-Nayestanak et al. (2012)]{kalantar12}  
Kalantar-Nayestanaki, N., Epelbaum, E., Messchendorp, J. G., \& Nogga, A. \ 2012, Rep. Prog. Phys., 75, 016301 

\bibitem[Kastaun \& Galeazzi (2015)]{Kastaun2015} Kastaun, W., \& Galeazzi, F.\ 2015, \prd, 91, 064027

\bibitem[Kievsky et al. (2018)]{kievsky18}
Kievsky, A., Viviani, M., Logoteta, D., Bombaci, I., \& Girlanda, L. \ 2018 Phys. Rev. Lett., 121, 072701

\bibitem[Kiuchi et al. (2014)]{kiuchi14} 
 Kiuchi, K., Kyutoku, K., Sekiguchi, Y., Shibata,  M. \& Wada, T. \ 2014 Phys. Rev. D, 90, 041502 

\bibitem[Lattimer \& Swesty (1991)]{LS} Lattimer, J., \& Swesty D. F. \ 1991, Nucl. Phys. A, 535, 331

\bibitem[Lattimer \& Prakash (2016)]{Latt-Prak-2016} 
Lattimer, J., \& Prakash, M. \ 2016, Physics Reports, 621, 127

\bibitem[Li et al. (2008)]{Li2008}  
Li, Z. H., Lombardo, U.,  Schulze, H.-J., \&  Zuo, W. \ 2008, Phys. Rev. C, 77, 034316

\bibitem[Li et al. (2006)]{ZHLi06}    
Li, Z. H., Lombardo, U., Schulze, H.-J., Zuo, W., Chen, L. W., \& Ma, H. R. \ 2006, Phys. Rev. C, 74, 047304  

\bibitem[Li \& Schulze (2008)]{li-schu_08}  Li, Z. H. \& Schulze, H.-J. \ 2008, Phys. Rev. C , 78, 028801 

\bibitem[Liebend{\"o}rfer et al.(2001)]{agile_paper} Liebend{\"o}rfer, M., Mezzacappa, A., \& Thielemann, F.-K.\ 2001, Phys. Rev. D, 63, 104003

\bibitem[Liebend{\"o}rfer et al.(2009)]{idsa_paper} Liebend{\"o}rfer, M., Whitehouse, S.~C., \& Fischer, T.\ 2009, \apj, 698, 1174

\bibitem[Logoteta et al. (2012)]{log2012}  
Logoteta, D., Bombaci, I., Provid{\^e}ncia, C., \& Vida\~na, I. \ 2012, Phys. Rev. D, 85, 023003


\bibitem[Logoteta et al. (2016b)]{logoteta16b} 
Logoteta, D., Bombaci, I., \& Kievsky, A. \ 2016b, Phys Rev. C, 94, 064001  

\bibitem[Logoteta et al. (2019)]{logoteta_nny} 
Logoteta, D., Vida\~na, I. \& Bombaci, I. \ 2019,  European Physical Journal A, 55, 207


\bibitem[Mezzacappa \& Bruenn(1993)]{Mezzacappa1993} Mezzacappa, A., \& Bruenn, S.~W.\ 1993, \apj, 405, 637


\bibitem[Miller et al. (2019)]{nicer1}
Miller, M. C. et al., \ 2019, Astrophys. J. Lett., 887, L24 

\bibitem[Most et al.(2020)]{Most2020} Most, E.~R., Jens Papenfort, L., Dexheimer, V., et al.\ 2020, European Physical Journal A, 56, 59


\bibitem[Oertel et al. (2017)]{oertel2017}  
Oertel, M., Hempel, M., Kl\"ahn, T., \& Typel, S. \ 2017, Rev. Mod. Phys., 89, 015007 

\bibitem[O' Connor et al. (2018)]{oconnor18} 
E. O' Connor et al., \ 2018, J. Phys. G: Nucl. Part. Phys., 45, 104001

\bibitem[Oppenheimer \& Volkoff (1939)]{OV39} 
Oppenheimer, J., \&  Volkoff, G. \ 1939, Phys. Rev, 55, 374 

\bibitem[Pan et al.(2019)]{Pan2019} Pan, K.-C., Mattes, C., O{\textquoteright}Connor, E.~P., et al.\ 2019, Journal of Physics G Nuclear Physics, 46, 014001


\bibitem[Perego et al. (2019)]{perego19} 
Perego, A., Bernuzzi, S. \& Radice, D. Eur. Phys. J. A \ 2019, 55, 124 

\bibitem[Piarulli et al. (2016)]{maria_local} 
Piarulli M., Girlanda, L., Schiavilla, R., Kievsky, A., Lovato, A., Marcucci, L. E., 
Pieper, S. C., Viviani, M. \&  Wiringa, R. B. \ 2016,  Phys. Rev. C, 94, 054007   

\bibitem[Piarulli et al. (2020)]{benchmark2020}
Piarulli M., Bombaci, I., Logoteta, D., Lovato A., \&  Wiringa, R. B. \ 2020,  Phys. Rev. C, 101, 045801  


\bibitem[Pons et al. (1999)]{pons99}
Pons, J. A., Reddy, S., Prakash, M., Lattimer, J. M., \& Miralles, J. A. \ 1999, ApJ, 513, 780


\bibitem[Prakash et al. (1997)]{prak97} Prakash M., Bombaci I., Prakash M., Ellis P. J.,  
Lattimer J. M., \& Knorren R. \ 1997,  Phys. Rep., 280, 1 

\bibitem[Radice et al.(2018a)]{Radice2018} Radice, D., Perego, A., Hotokezaka, K., et al.\ 2018, \apj, 869, 130

\bibitem[Radice et al.(2018b)]{Radice2018b} Radice, D., Perego, A., Bernuzzi, S., et al.\ 2018, \mnras, 481, 3670

\bibitem[Radice et al.(2020)]{Radice_review_2020} Radice, D., Bernuzzi, S., \& Perego, A.\ 2020, arXiv e-prints, arXiv:2002.03863


\bibitem[Rajaraman \& Bethe (1967)]{rajaraman-bethe67}
Rajaraman, R., \&  Bethe, H. A. \ 1967, Rev. Mod. Phys., 39, 745  

\bibitem[Rezzolla \& Takami (2016)]{RT2016} 
Rezzolla, L., \& Takami, K. \ 2016, Phys. Rev. D, 93, 124051 


\bibitem[Riley et al. (2019)]{nicer2}
Riley, T. E. et al., \ 2019, Astrophys. J. Lett., 887, L21 

\bibitem[Rios et al. (2006)]{rios06}
Rios A., Polls, A., Ramos, A., \& Muther, H., \ 2006 Phys. Rev. C 74, 054317 

\bibitem[Roberts et al.(2012)]{roberts2012} Roberts, L.~F., Shen, G., Cirigliano, V., et al.\ 2012, \prl, 108, 061103

\bibitem[Roberts \& Reddy(2017)]{roberts_PNS_2017} Roberts, L.~F., \& Reddy, S.\ 2017, Handbook of Supernovae, 1605

\bibitem[Schneider et al.(2019)]{Schneider2019} Schneider, A.~S., Roberts, L.~F., Ott, C.~D., et al.\ 2019, \prc, 100, 055802


\bibitem[Shen et al. (1998)]{shen98}
Shen, H., Toki, H., Oyamatsu, K., \& Sumiyoshi, K. \ 1998, Nucl. Phys. A, 637, 435

\bibitem[Shibata \& Hotokezaka(2019)]{Shibata_review_2019} Shibata, M., \& Hotokezaka, K.\ 2019, Annual Review of Nuclear and Particle Science, 69, 41

\bibitem[Shibata et al. (2005)]{shibata2005} 
Shibata, M., Taniguchi, K., \& Uryu, K. \ 2005, Phys. Rev. D, 71, 084021 


\bibitem[Schneider et al. (2017)]{SRO}  
Schneider, A. S., Roberts, L. F., \& Ott, C. D. \ 2017, Phys. Rev. C 96, 065802 


\bibitem[Song et al. (1998)]{song98}  
Song H. Q.,  Baldo M., Giansiracusa G., \&  Lombardo U. \ 1998 Phys. Rev. Lett., 81, 1584 


\bibitem[Steiner et al. (2013c)]{SFHO} 
Steiner, A. W., Hempel, M., \&  Fisher, T. \ 2013, Astrophys. J., 774, 17 

\bibitem[Strobel et al. (1999)]{stro}
Strobel, K., Schaab, C., \& Weigel, M. K.\ 1999, A\&A, 350, 497

\bibitem[Sugahara \& Toki (1994)]{TM1} 
Sugahara, Y., \& Toki, H. \ 1994 Nucl. Phys. A, 579, 557 


\bibitem[Timmes \& Arnett (1999)]{timmes}
Timmes, F. X., \& Arnett, D. \ 1999, ApJS, 125, 277

\bibitem[Togashi et al. (2017)]{togashi17} 
Togashi, H., Nakazato, K., Takehara, Y., Yamamuro, S., Suzuki, H., \& Takano, M.  \ 2017,  
Nucl. Phys. A, 961, 78  

\bibitem[Tolman (1934)]{Tol34} 
Tolman, R. C. \ 1934, Proc. Nat. Acad. Sci. (USA), 20, 3 
  
\bibitem[Typel et al. (2010)]{DD2} 
Typel, S., Ropke, G., Klahn, T., Blaschke, D., \& Wolter, H. H.  \ 2010, Phys. Rev. C, 81, 015803 

\bibitem[Vida\~na \& Bombaci (2002)]{vb02}   
Vida\~na, I., \& Bombaci, I. \ 2002, Phys. Rev. C, 66, 045801 

\bibitem[Vida\~na et al. (2011)]{hyp2}  
Vida\~na, I. Logoteta, D., Provid{\^e}ncia, C., Polls, A., \& Bombaci, I. \ 2011, EPL (Europhysics Letters), 94, 11002  

 
\bibitem[Wellenhofer  et al. (2016)]{WHK16}   Wellenhofer, C., Holt, J. W., \& Kaiser, N.  \ 2016, Phys. Rev. C, 93, 055802

\bibitem[Wellenhofer  et al. (2015)]{WHK_2015}   Wellenhofer, C., Holt, J. W., \& Kaiser, N.  
\ 2015, Phys. Rev. C, 92, 015801 

\bibitem[Wiringa et al. (1988)]{wff88}  
Wiringa, R.B., Fiks, V., \& Fabrocini, A.\ 1998, Phys. Rev. C, 38, 1010 

\bibitem[Woosley \& Heger(2007)]{WoosleyHeger2007} Woosley, S.~E., \& Heger, A.\ 2007, \physrep, 442, 269

\bibitem[Zappa et al.(2018)]{Zappa2018} Zappa, F., Bernuzzi, S., Radice, D., et al.\ 2018, \prl, 120, 111101

\bibitem[Zuo et al. (2014)]{zuo14} Zuo, W.,  Bombaci, I., \&  Lombardo, U. \ 2014, Eur. Phys. J. A, 50, 12  

\bibitem[Zuo et al. (1999)]{zuo_M2} Zuo, W.,  Bombaci, I., \&  Lombardo, U. \ 1999, Phys. Rev. C., 60, 024605  
\bibitem[Zuo et al. (2006)]{zuo_FT} Zuo, W., Li, Z. H., Lombardo, U., Lu, G. C., \& Schulze H.-J. \ 2006, Phys. Rev. C., 73, 035208

\bibitem[Zuo et al. (2004)]{zuo_2004}  Zuo, W., Li, Z. H., Li, A., \&  Lu, G. C.  \ 2004, Phys. Rev. C., 69, 064001 



\end{thebibliography}
\end{document}